\documentclass[12pt]{article}
\usepackage[margin=2cm]{geometry}
\usepackage{amsfonts,amssymb,epsfig,amsmath}
\usepackage{slashed}
\usepackage{color}
\usepackage{hyperref}

\renewcommand{\text}[1]{#1}

\newcommand{\be}{\begin{equation}}
\newcommand{\ee}{\end{equation}}
\newcommand{\ben}{\begin{displaymath}}
\newcommand{\een}{\end{displaymath}}
\newcommand{\bea}{\begin{eqnarray}}
\newcommand{\eea}{\end{eqnarray}}
\newcommand{\bean}{\begin{eqnarray*}}
\newcommand{\eean}{\end{eqnarray*}}
\newcommand{\nn}{\nonumber \\}
\newcommand{\ba}{\begin{array}}
\newcommand{\ea}{\end{array}}
\newcommand{\bi}{\begin{itemize}}
\newcommand{\ei}{\end{itemize}}

\def\l{\lambda}

\def\g{\gamma}
\def\G{\Gamma}

\def\G{\Gamma}
\def\g{\gamma}
\def\e{\epsilon}

\def\e{\epsilon}

\DeclareMathOperator{\re}{Re}
\DeclareMathOperator{\im}{Im}

\DeclareMathOperator{\vol}{vol}

\newcommand{\dd}{\mathrm{d}}
\newcommand{\DD}{\mathrm{D}}

\begin{document}
\begin{titlepage}

\vfill
\begin{flushright}
DMUS-MP-15-10 \\
YITP-SB-15-27 \\
CAS-KITPC/ITP-452
\end{flushright}

\vfill

\begin{center}
   \baselineskip=16pt
   {\Large \bf 3D supergravity from wrapped M5-branes}

   \vskip 2cm
     Parinya Karndumri$^a$, Eoin \'O Colg\'ain$^{b, c, d, e}$

       \vskip .6cm
             \begin{small}
             \textit{$^a$ Department of Physics, Faculty
of Science, Chulalongkorn University, 254 Phayathai Road, Pathumwan,
Bangkok 10330, THAILAND}
                 \vspace{3mm}
                 
                 \textit{$^b$ C.N.Yang Institute for Theoretical Physics, SUNY Stony Brook, NY 11794-3840, USA}
                 \vspace{3mm}

                  \textit{$^c$ Department of Mathematics, University of Surrey, Guildford GU2 7XH, UK}
                 \vspace{3mm}

                  \textit{$^d$ Korea Institute for Advanced Study, 85 Hoegiro, Seoul 130-722, KOREA}
                 \vspace{3mm}

                  \textit{$^e$ Kavli Institute for Theoretical Physics China, Institute for Theoretical Physics, \\ Chinese Academy of Sciences, Beijing 100190, CHINA}
                 \vspace{3mm}

             \end{small}
\end{center}

\vfill \begin{center} \textbf{Abstract}\end{center} \begin{quote}
Through consistent Kaluza-Klein reduction, we construct 3D $\mathcal{N} =2$ gauged supergravities corresponding to twisted compactifications of M5-branes on a product of constant curvature Riemann surfaces, including K\"{a}hler-Einstein four-manifolds.  We extend the reduction to fermionic supersymmetry variations in order to determine the 3D Killing spinor equations and classify all timelike supersymmetric solutions. As a by-product, we identify an infinite class of new supersymmetric warped $AdS_3$ (G\"odel) and warped $d S_3$ solutions. Moreover, we show that the superpotential $T$ encodes the central charge and R symmetry of the dual $\mathcal{N} = (0,2)$ SCFTs in the large $N$ limit. We demonstrate that the R symmetry matches the canonical $U(1)$ isometry from existing classifications of supersymmetric $AdS_3$ solutions to 11D supergravity with $\mathcal{N} = (0,2)$ supersymmetry. 

\end{quote}

\end{titlepage}


\section{Introduction}
3D gravity has no propagating degrees of freedom. Remarkably, despite being topological and admitting a Chern-Simons formulation \cite{Achucarro:1987vz, Witten:1988hc}, the theory is rich enough in $AdS_3$ to have black hole solutions \cite{Banados:1992wn}. Moreover, a large class of higher-dimensional black holes possess near-horizons with $AdS_3$ factors and it is striking that the geometrical Bekenstein-Hawking entropy is encoded in the central charge of a 2D conformal field theory (CFT) \cite{Strominger:1996sh,Strominger:1997eq}. In fact, a decade prior to the AdS/CFT conjecture \cite{Maldacena:1997re}, it was already established that the asymptotic symmetry group of $AdS_3$ was generated by two copies of the Virasoro algebra \cite{Brown:1986nw}. Thus, gravity in $AdS_3$ can be said to define a CFT.

In special settings, for example, M5-branes wrapped on Calabi-Yau (CY$_3$) four-cycles,  it is possible to go beyond the leading Bekenstein-Hawking entropy and compare one-loop corrections \cite{Maldacena:1997de, Harvey:1998bx}. For $AdS_3$ geometries with arbitrary higher-derivative terms, the central charge may be determined by extremising the on-shell action \cite{Kraus:2005vz}, thereby generalising the Brown-Henneaux result. A considerable advantage of this approach is that it does not assume supersymmetry.

For supersymmetric $AdS_3$ near-horizons, there is a recognisable redundancy in extremising the on-shell action. In fact, given sufficient knowledge of the effective 3D supergravity, one can simply extremise the superpotential $T$ to localise the action on supersymmetric configurations. In fact, for holographic RG flows interpolating between $AdS_3$ vacua, it is well-known \cite{Freedman:1999gp, Berg:2001ty} that the inverse of $T$ plays the r\^ole of the monotonically decreasing Zamolodchikov $c$-function \cite{Zamolodchikov:1986gt}. Furthermore, for 2D QFTs with $\mathcal{N} = (0,2)$ supersymmetry, a setting where the $U(1)$ R symmetry is ambiguous - it is free to mix with other $U(1)$ flavour symmetries - there is a well-defined procedure, $c$-extremization \cite{Benini:2012cz, Benini:2013cda} \footnote{We now have both black hole \cite{Kraus:2005vz} and CFT $c$-extremization. Settings can be found, e.g. black string solutions \cite{Gibbons:1994vm}, where these procedures agree.} to determine the central charge and R symmetry at superconformal fixed-points exactly \footnote{See \cite{Bobev:2014jva, Bertolini:2014ela, Chen:2014efa, Nagasaki:2014xya} for related recent  work.}. Since $T$ shares the same components as the Maxwell Chern-Simons (CS) terms, which in turn are fixed by anomalies, $T$ also knows about the R symmetry in the large $N$ limit \cite{Karndumri:2013iqa, Karndumri:2013dca}. So for two-derivative supergravity, it makes sense to study $T$. It remains to be seen if a counterpart exists with higher-derivatives, one that would potentially provide a repackaging of a recent \textit{tour de force} calculation involving 5D supergravity with four-derivative terms \cite{Baggio:2014hua}.

In this paper we continue a program \cite{Karndumri:2013iqa, Karndumri:2013dca,Colgain:2010rg} of identifying 3D $\mathcal{N}=2$ gauged supergravities \cite{deWit:2003ja} corresponding to wrapped-brane geometries. We recall that $AdS_3$ geometries - alternatively, the vacua of 3D gauged supergravities - based on wrapped-branes were initially constructed in lower-dimensional supergravity, e. g. \cite{Maldacena:2000mw, Gauntlett:2000ng, Gauntlett:2001jj, Cucu:2003yk, Cucu:2003bm}, before being uplifted to higher dimensions using consistent Kaluza-Klein (KK) sphere reductions \cite{Cvetic:1999xp,Liu:1999ai, Cvetic:1999un} \footnote{Further Kaluza-Klein embeddings of the same theories include \cite{Jeong:2013jfc, Colgain:2014pha}.}. In this work, we consider 7D $U(1)^2$ gauged supergravity \cite{Liu:1999ai}, which we twist and KK reduce on a product of constant curvature genus $\frak{g}_i$ Riemann surfaces, $\Sigma_{\frak{g}_1} \times \Sigma_{\frak{g}_2}$, and recast the 3D effective theory in the natural language of 3D gauged supergravity.  While our work here does not exhaust the possibilities for M5-branes wrapped on four-cycles - we have omitted K\"ahler four-cycles in Hyper-K\"ahler manifolds and co-associative cycles in $G_2$-holonomy manifolds - the ansatz is rich enough to include K\"ahler-Einstein (KE$_4$) compactifications as a special case and still allow for mixing of the R symmetry.

Given the existence of the 3D theory, it is reasonable to enquire into its solutions, particularly the supersymmetric solutions where powerful techniques exist  \cite{Gauntlett:2002nw} to find closed expressions. Similar studies have appeared recently for ungauged \cite{Deger:2010rb, deBoer:2014iba,Deger:2015tra}, gauged \cite{Colgain:2015mta} and massive gravity \cite{Deger:2013yla, Alkac:2015lma} in 3D. As we will show for our 3D gauged supergravity, supersymmetric spacetimes are characterised by a 2D Riemann surface and differential equations for the warp factor, $D$, and the canonical scalars $W_I$.  As it turns out, the superpotential $T$ also determines all timelike supersymmetric solutions to the 3D gauged supergravity; the field strengths are given in terms of the scalars $W_I$ and derivatives of the superpotential $\partial_{W_I} T$,  while the supersymmetry equations are expressed in terms of $T$ and $\partial_{W_I} T$. At no point does the explicit expression for $T$ appear, suggesting that this is a universal result. Therefore, for any 3D $\mathcal{N} =2$ gauged where the $U(1)$ R symmetry is gauged, once one determines $T$, one can simply write down the equations for all timelike supersymmetric solutions \footnote{See  \cite{Colgain:2015mta} for results on null solutions.}. To the extent of our knowledge, no classification of the supersymmeric solutions of 7D $U(1)^2$ gauged supergravity exists \footnote{See \cite{Cariglia:2004qi} for a classification of minimal gauged supergravity in 7D.}, in contrast to 5D $U(1)^3$ supergravity \cite{Gutowski:2004yv}. It is expected that our results in section \ref{sec:timelike} will serve as a consistency check for any future classification.

One interesting feature of the solutions we find is that, depending on where one is in parameter space, $AdS_3$ may not be the only supersymmetric critical point, i. e. solution  with constant $W_I$. Indeed, the theory typically admits new flux-supported geometries, corresponding to warped $AdS_3$ (G\"{o}del) \cite{Godel, Reboucas:1982hn} and warped $d S_3$, with characteristic closed timelike curves (CTCs). It is noteworthy that the warped and unwarped solutions appear at \textit{different} values of the scalars, so that the one-to-one map between charges in $AdS_3$ and warped $AdS_3$ identified in \cite{Compere:2014bia}, providing the basis for two copies of the Virasoro algebra in warped $AdS_3$, cannot apply, since the scalars are now dynamical. It would be interesting to extend the analysis of ref. \cite{Compere:2014bia} (also ref. \cite{Compere:2007in}) to 3D theories with scalar potentials to see whether the inverse of $T$, as suggested in \cite{Colgain:2015ela}, also encodes the central charge for warped $AdS_3$ solutions.

The structure of this paper is as follows. In section \ref{sec:KKred} we present the details of the twisted compactification from 7D supergravity and the rewriting of the bosonic action in terms of the canonical form for a 3D supergravity \cite{deWit:2003ja}, namely a non-linear sigma-model coupled to gravity. We identify the superpotential $T$ for the theory and the four complex scalars filling out the K\"{a}hler target space, $[SU(1,1)/U(1)]^4$, noting that this is, up to factors, the same target space that arises from KK reductions from IIB supergravity on both $S^1 \times \Sigma_{\frak{g}} \times \textrm{KE}_4$ \cite{Karndumri:2013dca} and M-theory on $S^2 \times \textrm{CY}_3$ \cite{Colgain:2010rg}.  In section \ref{sec:timelike}, we present closed expressions for all timelike supersymmetric solutions to the 3D supergravity through reduction of the supersymmetry variations from 7D \cite{Pernici:1984xx}, thus mirroring the analysis of ref. \cite{Colgain:2015mta}. We next employ standard Killing spinor bilinear techniques to derive the differential conditions on the spacetime. In section \ref{sec:c_ext} we review $c$-extremization for M5-branes wrapped on a product of Riemann surfaces \cite{Benini:2013cda}. From the extremal value of $T$, we show that one can read off the central charge and R symmetry in the large $N$ limit. Finally, in section \ref{sec:11Duplift}, we uplift the supersymmetric $AdS_3$ vacua to 11D and show that it fits into a known class of supersymmetric solutions \cite{Figueras:2007cn}.  

\section{3D gauged supergravity}
\label{sec:KKred}
\setcounter{equation}{0}
In this section we identify the bosonic sector of the Abelian 3D $\mathcal{N} =2$ gauged supergravity that arises when one truncates and consistently reduces 7D $SO(5)$ gauged supergravity \cite{Pernici:1984xx} on a product of Riemann surfaces. To support the claim that the effective 3D action is indeed a supergravity, we will demonstrate that the structure of the theory corresponds to the expected form of an $\mathcal{N} =2$ gauged supergravity \cite{deWit:2003ja}. To do this, we rewrite the action as a non-linear sigma-model with a K\"{a}hler target space. Combining our reduction ansatz with those of refs. \cite{Nastase:1999cb, Nastase:1999kf}, this provides an embedding of the 3D theory directly in 11D supergravity.

We begin by recalling the bosonic sector of maximally supersymmetric $SO(5)$ gauged supergravity in 7D \cite{Pernici:1984xx}; the theory comprises a metric, $SO(5)$ Yang-Mills fields $A^{ij}$, $i, j =1, \dots ,5$, five three-forms, $S^i$, transforming in the $\mathbf{5}$ of $SO(5)$ and 14 scalars parametrising the coset $SL(5, \mathbb{R})/SO(5)$ through the unimodular symmetric matrix $T_{ij}$. The bosonic action for this theory may be expressed as
\bea
\mathcal{L}_7 &=& R * \mathbf{1} - \frac{1}{4} T_{ij}^{-1} * \DD T_{jk} \wedge T^{-1}_{kl} \DD T_{li} - \frac{1}{4} T_{ik}^{-1} T^{-1}_{jl} * F^{ij} \wedge F^{kl} - \frac{1}{2} T_{ij} * S^i \wedge S^j \nn &+& \frac{1}{2g_7} S^i \wedge \DD S^i - \frac{1}{8 g_7} \epsilon_{j_1 \dots j_5} S^{j_1} \wedge F^{j_3 j_4} \wedge F^{j_4 j_5} - V * \mathbf{1} + \frac{1}{8 g_7} (2 \Omega_{5}[A]-\Omega_3[A]) , \nonumber
\eea
where $\Omega_{5}[A]$ and $\Omega_{3}[A]$ denote Chern-Simons forms for the gauge fields $A^{ij}$ \footnote{Taking into account the rescaling $A^{ij}_{\textrm{here}} = 2 A^{ij}_{\textrm{there}}$, they may be written in our notation as \cite{Pernici:1984xx}
\bea
\Omega_{3}[A] &=& \tfrac{1}{16} \epsilon^{\alpha \beta \gamma \delta \epsilon \eta \zeta} \textrm{Tr} (A_{\alpha} F_{\beta \gamma} - \tfrac{1}{3} A_{\alpha} A_{\beta} A_{\gamma}) \textrm{Tr}(F_{\delta \epsilon} F_{\eta \zeta}), \nn
\Omega_{5}[A] &=& \tfrac{1}{16} \epsilon^{\alpha \beta \gamma \delta \epsilon \eta \zeta} \textrm{Tr} \biggl(A_{\alpha} F_{\beta \gamma} F_{\delta \epsilon} F_{\eta \zeta} - \tfrac{2}{5} A_{\alpha} A_{\beta} A_{\gamma} F_{\delta \epsilon} F_{\eta \zeta} - \tfrac{1}{5} A_{\alpha} A_{\beta} F_{\gamma \delta} A_{\epsilon} F_{\eta \zeta}\nn &+& \tfrac{1}{5} A_{\alpha} A_{\beta} A_{\gamma} A_{\delta} A_{\epsilon} F_{\eta \zeta} - \tfrac{1}{35} A_{\alpha} A_{\beta} A_{\gamma} A_{\delta} A_{\epsilon} A_{\eta} A_{\zeta} \biggr).
\eea },
and we have defined
\bea
\DD T_{ij} &\equiv& \dd T_{ij} + g_7 A^{ik} T_{kj} + g A^{jk} T_{ik}, \nn
\DD S^i &\equiv& \dd S^i + g_7 A^{ij} \wedge S^j, \nn
F^{ij} &\equiv& \dd A^{ij} + g_7 A^{ik} \wedge A^{kj}.
\eea
The potential is given by
\be
V = \frac{g_7^2}{2} \left(2 T_{ij} T_{ij} - (T_{ii})^2 \right),
\ee
where $g_7$ is the gauge coupling.

Given the full $SO(5)$ theory, we can perform a group-theoretic truncation to 7D $U(1)^2$ gauged supergravity \cite{Liu:1999ai} (see also \cite{Nastase:2000tu}) by retaining two scalars, $\lambda_I$ \footnote{This theory is a further truncation of the $SO(4) \simeq SU(2) \times SU(2)$ theory with topological mass \cite{Karndumri:2014pla}.},
\be
\label{Tij}
T_{ij} = \textrm{diag}(e^{2 \lambda_1}, e^{2 \lambda_1}, e^{2 \lambda_2}, e^{2 \lambda_2}, e^{-4 \lambda_1 -4 \lambda_2} ),
\ee
two gauge fields $F^{12} = 2 \tilde{F}^{1}$, $F^{34} = 2 \tilde{F}^2$ and a three-form, $S^5 = 2 \sqrt{3} g_7 C$ \footnote{Note we are now using the scalings for the fields as they originally appeared in \cite{Pernici:1984xx}. The gauge couplings are now simply related through $g_7=m$.}, with all other fields set to zero.  This leads to the simpler bosonic action \cite{Liu:1999ai}
\bea
\mathcal{L}_7 &=& R * \mathbf{1} - 5 * \dd (\lambda_1 + \lambda_2) \wedge \dd (\lambda_1 + \lambda_2) -  * \dd (\lambda_1 - \lambda_2) \wedge  \dd (\lambda_1 - \lambda_2)\nn
&-& \sum_{i=1}^2 2 e^{-4 \lambda_i} * F^{i} \wedge F^i -  6 g_7^2 e^{-4 \lambda_1 -4 \lambda_2}  * C \wedge C  + 6 g_7 C \wedge \dd C- V * \mathbf{1} \nn
&-& 8 \sqrt{3} C \wedge F^1 \wedge F^2 + \frac{4}{g_7} \left( A^1 \wedge F^1 \wedge F^2 \wedge F^2 + A^2 \wedge F^2 \wedge F^1 \wedge F^1 \right),
\eea
where we have dropped tildes on $A^I$, since there is hopefully now no confusion regarding the origin of the truncated gauge fields. The potential also simplifies accordingly,
\be
V = \frac{g_7^2}{2} \left( - 8 e^{2 \l_1 + 2 \l_2} - 4 e^{-2 \l_1 - 4 \l_2} - 4 e^{-4 \l_1 - 2 \l_2} + e^{-8 \l_1 - 8 \l_2} \right).
\ee

To perform a reduction to 3D we now have options. Firstly, if we consider a reduction on the product of two genus $\frak{g}_i$ Riemann surfaces, $\Sigma_{\frak{g}_1} \times \Sigma_{\frak{g}_2}$, we could firstly reduce on one Riemann surface, thus making contact with the results of ref. \cite{ Szepietowski:2012tb, Baggio:2014hua} in 5D. However, experience suggests  \cite{Karndumri:2013iqa} that the reduction is suitably simple that it can be performed at the level of the action, so we opt to plough ahead and focus on the 3D theory.  We have independently checked that the reduction may be performed at the level of the equations of motion (EOMs), so it is by definition consistent.  Furthermore, by choosing the curvatures of the Riemann surfaces to be the same, one can replace $\Sigma_{\frak{g}_1} \times \Sigma_{\frak{g}_2}$ with a more general 4D K\"ahler-Einstein manifold, KE$_4$. Therefore, our analysis is expected to cover dimensional reduction on KE$_4$, simply through further truncation. This will be evident when we come to compare with the results in $c$-extremization \cite{Benini:2013cda} in a later section.

To perform the reduction on $\Sigma_{\frak{g}_1} \times \Sigma_{\frak{g}_2}$, we employ the spacetime ansatz
\be
\dd s^2_7 =  e^{-4(\lambda_3+\lambda_4)} g_{\mu \nu} \dd x^{\mu} \dd x^{\nu} +
e^{2 \lambda_3} \dd s^2 (\Sigma_{\frak{g}_1}) + e^{2 \lambda_4}
\dd s^2(\Sigma_{\frak{g}_2}),\label{7D_metric}
\ee
where $\lambda_I$, $I = 3, 4$ are scalar warp-factors for Riemann surfaces with constant curvature $\kappa_i,~ i =1, 2$ and $g_{\mu \nu}$ is the 3D metric in Einstein frame. The warp factors conspire to bring us to Einstein frame upon reduction. We also adopt an accompanying ansatz for the fluxes
\bea
\label{fluxes}
F^1&=& \frac{1}{2} G^1 - \frac{a_1}{4} \vol (\Sigma_{\frak{g}_1} )- \frac{a_2}{4} \vol(\Sigma_{\frak{g}_2}), \nn
F^2 &=& \frac{1}{2} G^2 - \frac{b_1}{4} \vol (\Sigma_{\frak{g}_1}) - \frac{b_2}{4} \vol(\Sigma_{\frak{g}_2}), \nn
C &=& \frac{\rho}{3!} \sqrt{-g} \epsilon_{\mu \nu \rho} \dd x^{\mu \nu \rho}+ \frac{1}{2} c_1\wedge
\vol (\Sigma_{\frak{g}_1})+ \frac{1}{2} c_2\wedge \vol (\Sigma_{\frak{g}_2}).
\eea

This introduces additional gauge fields, $B^I$, with field strengths $G^I = \dd B^I$, and one-forms $c_I$, which will be rewritten as field strengths later so that they conform to the canonical structure of 3D gauged supergravity \cite{deWit:2003ja}. The normalisation of the constant twist parameters, $a_i, b_i$ has been chosen to facilitate direct comparison with \cite{Benini:2013cda} and the factor $\rho$ is fixed by the equation of motion for $C$ to be
\be
\label{rho}
\rho = \frac{1}{8\sqrt{3} g_7^2} (a_1 b_2 + b_1 a_2) e^{4 (\l_1 + \l_2) -8 (\lambda_3+\lambda_4)}.
\ee

Performing the reduction either at the level of the action or the EOMs, we arrive at the 3D action:
\bea
\label{action}
\mathcal{L}_{3} &=&   R *_3 \mathbf{1}- \sum_{I=1}^2 \biggl[ 5 *_3 \dd (\lambda_{2 I-1}+\lambda_{2 I}) \wedge \dd (\lambda_{2 I-1}+\lambda_{2I}) + *_3 \dd (\lambda_{2 I-1}-\lambda_{2I}) \wedge \dd (\lambda_{2 I-1}-\lambda_{2I}) \biggr] \nn &-& \frac{3}{2} g_7^2 e^{-4( \l_1 + \l_2)} \sum_{I=1}^2 e^{-4 \lambda_{2 + I}} *_3 c_{I} \wedge c_I - \frac{1}{2} e^{4(\lambda_3+\lambda_4)}\sum_{I=1}^2e^{-4\lambda_I} *_3 G^I \wedge  G^{I} - V_3 *_3 \mathbf{1} +\mathcal{L}_{CS}, \nonumber
\eea
where the 3D potential is now
\bea
V_3 &=& \frac{1}{2} g_7^2 e^{-4 (\lambda_3 + \lambda_4)} \biggl[ - 8 e^{2 \l_1 + 2 \l_2} - 4 e^{-2 \l_1 - 4 \l_2} - 4 e^{-4 \l_1 - 2 \l_2} + e^{-8 \l_1 - 8 \l_2} \biggr]  \nn
&+& \frac{1}{8} e^{-8 \lambda_3-4 \lambda_4}\left(a_1^2e^{-4\lambda_1}+b_1^2e^{-4\lambda_2}\right)+ \frac{1}{8} e^{-4 \lambda_3-8 \lambda_4}\left(a_2^2e^{-4\lambda_1}+b_2^2e^{-4\lambda_2}\right) \nn
&-& 2e^{-4( \lambda_3+ \lambda_4)}\left(\kappa_1e^{-2 \lambda_3}+\kappa_2e^{-2 \lambda_4}\right) +
\frac{1}{32 g_7^2} (a_1 b_2 + b_1 a_2)^2 e^{4(\l_1 + \l_2)}
e^{-8(\lambda_3+\lambda_4)},\label{V3_from7D}   \nonumber \eea
and the CS term
becomes \bea \mathcal{L}_{\textrm{CS}} &=&
\label{CS_action}
\frac{1}{4 g_7} \left[b_1b_2 B^{1}\wedge
G^{1}+(a_1b_2+a_2 b_1)B^{1}\wedge G^{2}+ a_1a_2B^{2}\wedge
G^{2}\right.\nonumber
\\& &\left.+(a_2 b_1+a_1b_2)B^{2}\wedge G^{1}\right]+ \frac{3}{2} g_7 (c_1\wedge \dd c_2+c_2\wedge
\dd c_1)\nonumber \\
&+& \frac{\sqrt{3}}{2} \left[c_1\wedge
\left(b_2G^{1}+a_2G^{2}\right)+c_2\wedge
\left(b_1G^{1}+a_1G^{2}\right)\right]. \eea

To recast the action in the accustomed form of a non-linear sigma-model coupled to supergravity, we normalise and diagonalise the scalar kinetic terms through the redefinitions:
\bea
\label{scalar_redef}
W_1 &=& - 2 (\l_1 - \l_3
-\l_4), \quad W_2 = - 2(\l_2 - \l_3 - \l_4), \nn W_3 &=& 2 (\l_1 + \l_2 +\l_4),
\quad W_4 = 2(\l_1 + \l_2 +\l_3). \label{Wi7D} \eea
With these redefintions, the potential may be written as
\be
\label{V_T}
V_3 = -8 T^2 + 8 \sum_{I=1}^4 (\partial_{W_I} T)^2,
\ee
where we have introduced the superpotential $T$:
\bea
T &=&  \frac{g_7}{4} \left( 2 e^{-W_1} + 2 e^{-W_2} + e^{-W_3 -W_4} \right) - \frac{1}{16 g_7} (a_1 b_2 + b_1 a_2) e^{-W_1 -W_2} \nn
&-& \frac{1}{8} \left( a_1 e^{-W_2 - W_4} + b_1 e^{-W_1 -W_4} + a_2 e^{-W_2 -W_3} + b_2 e^{-W_1 -W_3} \right).
\eea
It can be checked that $T$ recovers the correct potential provided the curvature of the Riemann surfaces is related to the twist parameters through the following supersymmetry condition
\be
\label{susy_cond}
\kappa_i = - \frac{g_7}{2} (a_i + b_i).
\ee
From (\ref{V_T}) it is clear that the critical points of $V_3$ correspond to $\partial_{W_I} T = 0$. As we show in the appendix, the same expression for $T$ also appears in the dimensional reduction of the fermionic supersymmetry conditions from 7D, thus providing further confirmation that we have reduced the theory correctly. Using the results in the appendix, it is easy to show that solving the Killing spinor equation to find $AdS_3$ vacua is equivalent to extremising the superpotential.

The condition (\ref{susy_cond}) guarantees that the lower-dimensional theory is indeed a gauged supergravity, one with $\mathcal{N} =2$ supersymmetry. Choosing $g_7=2$, which leads to the canonical normalisation for $AdS_7 \times S^4$ so that the radius of the original $AdS_7$ vacuum is unity, (\ref{susy_cond}) corresponds with the supersymmetry conditions presented in \cite{Benini:2013cda}.

To back up our claim that the theory corresponds to an $\mathcal{N}=2$ gauged supergravity, we need to demonstrate that there is a K\"{a}hler scalar manifold. To show this is indeed the case, we record the following equations of motion that follow from the 3D action
\bea
\label{c1} \dd c_1 &=& g_7 \, e^{-2 W_3} *_3 c_2 - \frac{1}{2 \sqrt{3} g_7} (b_1 G^1 + a_1 G^2), \\
\label{c2} \dd c_2 &=& g_7  \, e^{-2 W_4} *_3 c_1 - \frac{1}{2 \sqrt{3} g_7} (b_2 G^1 + a_2 G^2),
\eea
\bea
\label{G1} \dd \left( e^{2 W_1} *_3 G^1 \right) &=& \frac{\sqrt{3}}{2} (b_1 \dd c_2 + b_2 \dd c_1) \nn && \phantom{xxxxxxxx} + \frac{1}{2g_7} \left[  b_1 b_2 G^1 + (a_1 b_2 + a_2 b_1) G^2 \right], \\
\label{G2} \dd \left( e^{2 W_2} *_3 G^2 \right) &=& \frac{\sqrt{3}}{2} (a_1 \dd c_2 + a_2 \dd c_1) \nn && \phantom{xxxxxxxx} + \frac{1}{2g_7} \left[  a_1 a_2 G^2 + (a_1 b_2 + a_2 b_1) G^1 \right].
\eea
An observation that we can make at this point is that one cannot truncate out $c_I$ without setting $G^I = 0$ for generic twists, in which case one recovers the ansatz of \cite{Benini:2013cda}. So, if we plan on retaining $G^I$, then we are forced to also incorporate $c_I$.  We now introduce scalars $Y_I$ through the following covariant derivatives
\bea
\label{cov_diff1} e^{2 W_1} *_3 G^1 =  \DD Y_1 &\equiv& \dd Y_1 + \frac{1}{2} ( b_2 A^1 + b_1 A^2) + \frac{1}{4g_7} (a_1 b_2 + a_2 b_1) B^2, \\
\label{cov_diff2} e^{2 W_2} *_3 G^2 = \DD Y_2 &\equiv& \dd Y_2 + \frac{1}{2} ( a_2 A^1 + a_1 A^2) +  \frac{1}{4g_7} (a_1 b_2 + a_2 b_1) B^1, \\
\label{cov_diff3} - \sqrt{3} g_7 c_2 = \DD Y_3 &=& \dd Y_3 + \frac{1}{2} (b_2 B^1 + a_2 B^2) - g_7 A^2, \\
\label{cov_diff4} - \sqrt{3} g_7 c_1 = \DD Y_4 &=& \dd Y_4 + \frac{1}{2} (b_1 B^1 + a_1 B^2) - g_7 A^1.
\eea
This ensures that (\ref{c1})-(\ref{G2}) are trivially satisfied. It is worth recording that the derivative (\ref{c1}) and (\ref{c2}) imply that $ e^{-2 W_3} * c_2 = \dd A$, where $A$ is an arbitrary one-form. The precise relationship can be fixed by comparing with (\ref{cov_diff3}) and (\ref{cov_diff4}), resulting in
\be
\label{c_dual}
e^{-2 W_4} *_3 c_1 = \frac{1}{\sqrt{3} g_7} F^2, \quad e^{-2 W_3} *_3 c_2 = \frac{1}{\sqrt{3} g_7} F^1,
\ee
where $F^I = \dd A^I$.

Once the scalars $Y_I$ are introduced, we can rewrite the kinetic terms as
\be
\mathcal{L}_{\textrm{scalar}} = - \frac{1}{2} \sum_{I=1}^4 \left[ *_3 \dd W_I \wedge \dd W_I + e^{-2 W_I}  *_3 \DD Y_I \wedge \DD Y_I \right].
\ee
The CS terms consistent with (\ref{cov_diff1})-(\ref{cov_diff4}) are
\bea
\label{CS2}
\mathcal{L}_{\textrm{CS}} &=& g_7 A^1 \wedge F^2
- \frac{1}{4 g_7} (a_1 b_2 + a_2 b_1) B^1 \wedge G^2 \nn
&-& \frac{1}{2} A^1 \wedge (b_2 G^1 + a_2 G^2) - \frac{1}{2} A^2 \wedge ( b_1 G^1 + a_1 G^2).
\eea
We are free to then introduce complex coordinates $z_I = e^{W_I} + i Y_I$ so that the K\"{a}hler potential for the manifold is
\be
\label{Kahler_pot}
K =  -\sum_{I=1}^4 \log [ \Re (z_I)],
\ee
thus demonstrating that the K\"{a}hler manifold is $[SU(1,1)/U(1)]^4$. This confirms that the bosonic action is consistent with $\mathcal{N} =2$ gauged supergravity \cite{deWit:2003ja}.

Extrema of $T$ correspond to supersymmetric $AdS_3$ vacua. In our notation, these may be expressed explicitly as
\bea
\label{T_extrema}
e^{-W_1}&=&\frac{4  \left(a_1b_2+a_2b_1-a_1a_2\right) g_7^2}{a_2^2b_1^2+a_1^2b_2^2+a_1a_2b_1b_2},\qquad
e^{-W_2}=\frac{4\left(a_1b_2+a_2b_1-b_1b_2\right)g_7^2}{a_2^2b_1^2+a_1^2b_2^2+a_1a_2b_1b_2},\nonumber
\\
e^{-W_3}&=&\frac{2\left(a_1^2b_2+a_2b_1^2\right)g_7}{a_2^2b_1^2+a_1^2b_2^2+a_1a_2b_1b_2},\qquad
~~~~e^{-W_4}=\frac{2\left(a_2^2b_1+a_1b_2^2\right) g_7}{a_2^2b_1^2+a_1^2b_2^2+a_1a_2b_1b_2}.
\eea
Using the redefinitions $z_i = a_i -b_i$, (\ref{scalar_redef}) with $g_7=2$, it can be checked that these agree with the critical points of \cite{Benini:2013cda} once a flip in the sign of the scalars $\lambda_1$ and $\lambda_2$ is taken into account. We recall that these vacua were originally found by solving the Killing spinor equations  \cite{Benini:2013cda}, whereas here we have simply identified and extremised the superpotential of the effective 3D theory. We stress that the above expressions for supersymmetric $AdS_3$ critical points hold for generic $a_i, b_i$ with $\mathcal{N} = (0,2)$ supersymmetry. Special points in parameter space exist where supersymmetry is enhanced to $\mathcal{N} = (0,4)$ supersymmetry \footnote{This corresponds to setting either $F^{12} = 0$ or $F^{34} = 0$ in 7D notation, so we only need to impose half the projection conditions given in (\ref{red_projection}).}, where $a_1 = a_2 = 0$ or $b_1 = b_2 = 0$, however it can be verified from the superpotential that no extremum exists for these values. Similarly, when $a_1 = b_1 =0$, or $a_2 = b_2 = 0$, there is no flux to support an $AdS_3$ vacuum and as a consequence there is no solution.

There is one special case with an $AdS_3$ vacuum and enhanced supersymmetry, which may be found by setting $a_1 = b_2 = 0$, or $a_2 = b_1 = 0$. Here supersymmetry is enhanced to $\mathcal{N} = (2,2)$, a feature that can be seen from (\ref{red_projection}), since we need only impose two projection conditions, $\gamma_{12} \Gamma^{12} \epsilon = \epsilon, \gamma_{34} \Gamma^{34} \epsilon = \epsilon$, resulting in eight supersymmetries. Extremising the potential, we note that
\be
W_1 = W_2 = \log \left( \frac{a_2 b_1}{4 g_7^2} \right), ~~W_3 =  \log \left( \frac{a_2}{2g_7} \right), ~~W_4 = \log \left( \frac{b_1}{2g_7} \right), \ee
where we have assumed $a_2$ and $b_1$ are non-zero. Indeed, for $W_I \in \mathbb{R}$, we further infer that $a_2 > 0$ and $b_1 > 0$, which implies through (\ref{susy_cond}), with positive $g_7$, that in order to preserve supersymmetry we must consider compactification on a product of hyperbolic spaces.

Setting $g_7=2$, one can quickly identify the compactifications leading to real $AdS_3$ vacua. One notes that solutions only exist for  $H^2 \times \Sigma_{\frak{g}}$, or put differently, one of the Riemann surfaces should be hyperbolic. Choosing $\kappa_1 = -1$ and $\kappa_2 \in \{0, \pm 1\}$, we note the following constraints on the parameters:
\bea
\label{parameter_constraints}
H^2 \times T^2 &\quad& \{ a_2 > 0, ~a_1 < \frac{1}{3} \} \cup \{ a_2 < 0, ~a_1 > \frac{2}{3} \}, \nn
H^2 \times S^2 &\quad& \{a_2 > 0,~ \frac{a_2}{3 a_2 +1} > a_1> -\frac{a_2^2}{2 a_2 +1} \} \cup \{ a_2 < -1, ~\frac{2 a_2 +1}{3 a_2 +2} < a_1 < - \frac{a_2^2}{2 a_2 +1} \}, \nn
H^2 \times H^2 &\quad& \{ 3(a_1 + a_2) - 6 a_1 a_2 -1 > | a_1 + a_2 -1| \},
\eea
where we have eliminated $b_i$ through (\ref{susy_cond}). We have checked that these agree with the parameter constraints given in \cite{Benini:2013cda}. Identifying $a_1 = a_2=a$, $b_1 = b_2 = b$, one can show similarly that twisted compactifications on KE$_4$ only lead to real $AdS_3$ vacua when $ \kappa = - (a+ b) < 0$, so the space is negatively curved.

We have also checked that our results at leading order are consistent with the 5D analysis presented in \cite{Baggio:2014hua}, where subleading corrections to the geometry are considered. More precisely, one can check that $e^{W_3}$ corresponds to the lone 5D hyperscalar, which appears upon reduction from 7D, and that the scalars in the three vector multiplets are
\bea
X^{1} = e^{- \frac{2}{3} \l_4 + 2 \l_1}, \quad X^{2} = e^{- \frac{2}{3} \l_4 + 2 \l_2}, \quad X^3 = e^{\frac{4}{3} \l_4 - 2 \l_1-2 \l_2}.
\eea
In terms of the remaining $W_I$, we have
\be
e^{-W_1} = \frac{A_1}{a_2 a_3}, \quad e^{-W_2} = \frac{A_2}{a_1 a_3}, \quad e^{-W_4} = \frac{A_3}{a_1 a_2},
\ee
where $A_1 = ({g_5}/{2}) (-a_1 P_1 + a_2 P_2 + a_3 P_3)$ is expressed in terms of the 5D gauge coupling, $g_5$, the 5D twist parameters $a_i$ and the moment maps \cite{Baggio:2014hua}:
\be
g_5 P_1 = 2 m - \frac{p_2}{2} e^{-W_3}, \quad g_5 P_2 =  2 m- \frac{p_1}{2} e^{-W_3}, \quad g_5 P_3 = m e^{-W_3},
\ee
with 7D gauge coupling $m$. Similar expressions can be found for $A_2, A_3$. To make the notation of ref. \cite{Baggio:2014hua} consistent with our notation, one should employ the following redefinitions:
\bea
a_1 &\rightarrow& \frac{a_1}{2}, \quad a_2 \rightarrow \frac{b_1}{2}, \quad p_1 \rightarrow a_2, \quad p_2 \rightarrow b_2, \nn
a_3 &\rightarrow& \frac{1}{4m} (a_1 b_2 + a_2 b_1), \quad m \rightarrow g_7.
\eea
One can also check that one of the conditions arising from the vanishing of the 5D hyperino variation
\be
k_{I}^X X_{I} = 0
\ee
where $k_I^{X}$ denote Killing vector parameters associated to a quaternionic submanifold of the hyperk\"ahler manifold corresponding to the hypermultiplets, is recast in 3D into the condition that the 3D superpotential is independent of the hyperscalar $\partial_{W_3} T = 0$.

With an eye on the analysis in section \ref{sec:c_ext}, we record the value of $T$ at the extremum,
\be
T =\frac{\left(2a_1b_2+2a_2b_1-a_1a_2-b_1b_2\right) g_7^3}{a_1^2b_2^2+a_2^2b_1^2+a_1a_2b_1b_2}.
\ee
This in turn sets the $AdS_3$ radius, $\ell$, through $\ell = 1/(2 T)$, as can be seen from the scalar potential.

\subsection{Further Truncations}
It is clear from the earlier analysis that one can further consistently truncate our theory. For example, for the choice of parameters $a_1 = a_2 = a$ and $b_1 = b_2 = b$, which implies $\kappa_1 = \kappa_2$, one may consider the simplification $W_3 = W_4$, $A^1 = A^2$ and this gauged supergravity with $[SU(1,1)/U(1)]^3$ K\"ahler target space contains information about reductions on KE$_4$. More precisely, in addition to the complex scalars $z_1, z_2$, which are unaffected, we retain $z_3 = e^{W_3} + i Y_3$ and the K\"{a}hler potential for the target becomes
\be
K = - \log[ \Re (z_1) ] - \log [ \Re (z_2) ] - 2 \log [ \Re(z_3) ].
\ee
and the scalar potential may be expressed as
\bea
V_3 &=& - 8 T^2 + 8 [ (\partial_{W_1} T)^2 + (\partial_{W_2} T)^2 + \frac{1}{2} (\partial_{W_3} T)^2],
\eea
where the superpotential is now
\bea
T &=&  \frac{g}{4} \left( 2 e^{-W_1} + 2 e^{-W_2} + e^{-2 W_3} \right) - \frac{ab}{8 g}  e^{-W_1 -W_2} - \frac{1}{4} e^{-W_3} \left( a e^{-W_2} + b e^{-W_1} \right).
\eea
Solving $\partial_{W_I} T = 0$, $I = 1, 2, 3$, one recovers the supersymmetric $AdS_3$ values (\ref{T_extrema}) with the constrained parameters, as expected. It is a simple exercise to consider further truncations to $[SU(1,1)/U(1)]^2$ target manifolds by identifying $W_1 = W_2$, etc.

\section{All Timelike solutions}
\label{sec:timelike}
In this section, noting that the 3D gauged supergravity in section \ref{sec:KKred} is structurally the same as the $U(1)^3$ theory presented in \cite{Colgain:2015mta}, we derive the general solution to all timelike supersymmetric solutions. In both cases, the respective 3D gauged supergravities possess K\"ahler target space $[SU(1,1)/U(1)]^{n}, n \in \{3, 4\}$, so it may be anticipated that supersymmetric geometries are the same. We remark that it is straightorward to generalise our results to arbitary $n \in \mathbb{N}$ in analogy with known 5D classifications \cite{Gutowski:2004yv}. While our interest here is gauged supergravity with scalar potentials, we also note that ungauged supergravities in 3D were classified in \cite{Deger:2010rb, Deger:2015tra, deBoer:2014iba}.

Following ref. \cite{Colgain:2015mta}, supersymmetric timelike solutions for the 3D gauged supergravity presented in section \ref{sec:KKred} take the form
\bea
\label{susy_solution}
\dd s^2_3 &=& - ( \dd \tau + \rho)^2 + e^{2 D -K} ( \dd x_1^2 + \dd x_2^2), \nn
G^{I} &=& e^{-W_I} \left[ -4 \partial_{W_I} T \, e^{2 D-K} \dd x_1 \wedge \dd x_2 + ( \dd \tau + \rho) \wedge \dd W_I\right], \nn
F^{I} &=& e^{-W_{I+2}} \left[ -4 \partial_{W_{I+2}} T \, e^{2 D-K} \dd x_1 \wedge \dd x_2 + ( \dd \tau + \rho) \wedge \dd W_{I+2}\right], ~~I = 1, 2,
\eea
where repeated $I$ indices in $F^{I}, G^{I}$ are not summed, $(x_1, x_2)$ parametrise a Riemann surface, $\rho$ is a one-form connection on the Riemann surface satisfying
\be
\label{connection}
\dd \rho = 4 T e^{2D-K} \dd x_1 \wedge \dd x_2, \ee
$D$ is the breathing mode for the Riemann surface and $K$ is the K\"{a}hler potential (\ref{Kahler_pot}).  We observe that the expression for the field strengths ensures that the algebraic 3D Killing spinor equations presented in the appendix (\ref{algebraic}) are satisfied. When the gauge fields are zero we have full supersymmetry, so it is hopefully clear that non-zero field strengths imply the projection condition $\gamma^{12} \xi = i \xi$, thus breaking supersymmetry by one-half, leaving generically two supersymmetries.

To see that the one-form connection must satisfy (\ref{connection}), as in \cite{Colgain:2015mta}, we can introduce the vector spinor bilinear $P^0_a \equiv \bar{\xi} \gamma_{a} \xi$ and make use of the Killing spinor equation (\ref{KSE}),
\be
 \biggl[ \mathcal{D}_{a} + T \gamma_{a} + \frac{i}{8} \sum_{i=1}^2 \left( e^{W_i} \gamma_{a}^{~bc}  G^i_{bc} + e^{W_{i+2}}  \gamma_{a}^{~bc} F^{i}_{bc} \right) \biggr] \xi = 0,
\ee
to determine that $\dd P^0 = 4 T *_3 P^0$. Further defining the complex vector bilinear $(P^1 + i P^2)_a \equiv \bar{\xi}^c \gamma_{a} \xi$,  we can use the same technique to find the differential condition:
\bea
\label{diff_cond}
e^{-\frac{1}{2} K} \dd \left[ e^{\frac{1}{2} K} (P^1 + i P^2) \right] &=& g_7 \, (e^{-W_1}+e^{-W_2}) *_3 (P^1 + i P^2)\nn
&+& i \, g_7  (B^1 + B^2) \wedge (P^1 + i P^2).
\eea
In this equation we note that the LHS does not depend on the timelike Killing direction, while the RHS does. As a result, since $B^{I}$ generically have electric components, for consistency we require that $B^I$ takes the form
\be
B^{I} = e^{-W_I} ( \dd \tau + \rho) + \tilde{B}^I,~~I = 1, 2,
\ee
where $\tilde{B}^I$ is a one-form depending only on the coordinates of the Riemann surface, $x_1, x_2$. Furthermore, with this choice for gauge potential,  $B^I$ is now consistent with the field strength $G^I$ (\ref{susy_solution}). From the same equation, we can determine the equation for the warp factor $D$. We note that the form of the metric in (\ref{susy_solution}) is consistent with the choice $P^1 + i P^2 = e^{D-\frac{1}{2}K } ( \dd x_1 + i \dd x_2)$. Inserting this expression for the complex vector into the above differential condition, we find that $ g_7\, (\tilde{B}^1 + \tilde{B}^2) = *_2 \dd D$. Taking a further derivative, we find a second order equation:
\be
\label{D_eqn}
\nabla^2 D  = 4 \, g_7  \sum_{I=1}^2 \left(  e^{-W_I} \partial_{W_I} T  + e^{-W_I} T \right) e^{2 D -K},
\ee
which is exactly the same as in the $U(1)^3$ theory \cite{Colgain:2015mta}, modulo a different expression for the superpotential $T$ and an overall factor of the coupling $g_7$. This is in line with our expectations. At fixed $W_I$, this equation is nothing more than the Liouville equation $\nabla^2 D = - \mathcal{K} e^{2D}$, where $\mathcal{K}$ is the Gaussian curvature of the Riemann surface. As discussed further in \cite{ Colgain:2015mta, Colgain:2015ela}, at extrema of the superpotential, the Gaussian curvature is related to the $AdS_3$ radius, $\ell$ and the extremal value of $T$ in the following fashion,
\be
\frac{4}{e^{K}\, \mathcal{K}}  |_{\rm ext}  = \ell^2 = \frac{1}{4 T^2} |_{\rm ext}.
\ee

To extract supersymmetry conditions for the scalars we can use the expressions for the field strengths (\ref{susy_solution}) in the flux EOMs:
\bea
\label{scalar_eqn}
e^{-W_I} \nabla^2 e^{W_I}&=& 16 \biggl[ \sum_{J \neq I} \partial_{W_J} T \, \partial^2_{W_I W_J} T - T \, \partial_{W_I} T \biggr] e^{2 D -K}.
\eea
We clearly note the presence of the supersymmetric critical point, where $\partial_{W_I} T = 0$, where $W_I$ becomes constant. It is interesting that the explicit expression for the superpotential does not appear, so this is presumably a general result for all 3D gauged supergravities with target space $[SU(1,1)/U(1)]^n, ~n \in \mathbb{N}$.

One can show that these conditions along with (\ref{susy_solution}) imply the scalar EOMs and the Einstein equations. As explicitly shown in \cite{Colgain:2015mta}, the Einstein equation along the temporal direction is trivially satisfied, whereas the Einstein equation along the Riemann surface reduces to the equation
\be
\left[ 16 T^2 - 8 ( \partial_{W_I} T)^2 \right] e^{2 D-K} - \nabla^2 D - \frac{1}{2} \sum_{I=1}^4 e^{-W_I} \nabla^2 e^{W_I} = 0.
\ee
Using (\ref{D_eqn}) and (\ref{scalar_eqn}) one can show that this equation is satisfied, thus providing us with a valuable consistency check.

\subsection{Warped geometries}
We now have closed expressions for all timelike supersymmetric solutions to the 3D gauged supergravity that arise through a compactification from 7D on a product of Riemann surfaces or a K\"ahler-Einstein four-manifold. As we have shown, the task of finding new solutions reduces to solving (\ref{D_eqn}) and (\ref{scalar_eqn}). In this subsection, we will focus on the simplest class of solutions with constant $W_I$ and leave more involved, potentially numeric solutions, to future work. This will lead to new solutions and, as a further consistency check, the recovery of supersymmetric $AdS_3$ vacua highlighted in section \ref{sec:KKred}.

Our strategy is then to consider fixed-points where $\dd W_I = 0$. As a consequence, the RHS of (\ref{scalar_eqn}) must vanish. In contrast to the simpler gauged supergravity of wrapped D3-brane geometries \cite{Colgain:2015mta}, here it is difficult to find analytic expressions for the  scalars in terms of our parameters, $a_i, b_i$. As a result, we adopt different means; we impose the quantisation condition (\ref{quant}) from the outset, thereby imposing a grid of discrete solutions, before sampling various points. Throughout, we use the coupling $g_7=2$.

We recall that when $W_I$ is constant, (\ref{D_eqn}) reduces to the Liouville equation, $\nabla^2 D = - \mathcal{K} e^{2 D}$ on the Riemann surface. A simple single-centered solution takes the form
\be
e^{D} = \frac{2 \sqrt{|\mathcal{K}|}}{|\mathcal{K}| + \mathcal{K} r^2}, 
\ee
leading to the 3D solution
\bea
\dd s^2 &=& -\ell_1^2 \left( \dd \tau + \frac{r^2}{(1 + {\rm sgn}(\mathcal{K}) r^2)} \dd \varphi \right)^2 + \ell_2^2 \frac{(\dd r^2 + r^2 \dd \varphi^2)}{(1+ {\rm sgn}(\mathcal{K}) r^2)^2}
\eea
where
\be
\ell^2_1 = \frac{64 T^2}{e^{2 K} |\mathcal{K}|^2} , \quad \ell_2^2 = \frac{4}{e^{K} |\mathcal{K}|}.
\ee
When $\ell_1 = \ell_2$ and $\mathcal{K} < 0$,  we recover unwarped $AdS_3$  \footnote{It is well-known that the BTZ black hole \cite{Banados:1992wn} is a quotient of $AdS_3$. While locally BTZ possesses as many supersymmetries as $AdS_3$, globally the number of supersymmetries depends on the mass, $M$, and angular momentum, $J$. For extremal black holes, $J= M \ell$, one supersymmetry is preserved, for $M=0$  two, and $M=-1$ ($AdS_3$) four \cite{Coussaert:1993jp}.  }.

When $\mathcal{K} < 0$, the Riemann surface is hyperbolic, whereas for $\mathcal{K} > 0$, we encounter a sphere. This can be easily seen by employing the coordinate transformations, $r = \tanh ({\rho}/{2})$, and $r = \tan ( {\theta}/{2})$, respectively. As a consequence, we see that for $\mathcal{K} < 0$, we have $ 0 \leq r \leq 1$, whereas for $\mathcal{K} > 0$, the radial direction is simply bounded below by zero,  $0 \leq r$. Regardless of the sign of the Gaussian curvature, we recognise that CTCs appear where the signature of the $g_{\varphi \varphi}$ term in the metric changes sign from positive to negative. The geometry is therefore CTC-free in the range
\be
r \leq \frac{\ell_2}{\ell_1}.
\ee
For unwarped $AdS_3$ this range coincides with the range of $r$, so there are no CTCs. To see that uplifting the warped vacua will make no difference to the presence of CTCs, we remark that when $\alpha = 0$ in the uplifted geometry (\ref{11D_uplift}), $A^{\alpha}$ makes no contribution and the problem reduces to analysing the presence of CTCs in the 3D metric, which we have done above. It should be clear that 3D CTCs will persist in 11D.

\begin{table}[h]
\begin{center}
\begin{tabular}{|c|c|c|c|c|}\hline
$(a_1, a_2)$ & $(\frak{g}_1, \frak{g_2})$  & $\phantom{e^{\hat{W_x}}}$ $(e^{W_1}, e^{W_2}, e^{W_3}, e^{W_4})$ & $\mathcal{K}$ & $ \ell_2/\ell_1$ \\
\hline
\hline
(0, 1) & ($n+1, 1$) & $(0.0625, 0.0313, 0.2500, 0.2500)$ & -1.1250 & 1 \\
(0.25, 1) & ($2 n+1, 1$) & (0.1094, 0.0219, 0.2188, 0.1094) & -0.6891 & 1 \\
& &  (0.0046, 0.0044, 0.2310, 0.0326) & 0.0743 & 0.1575 \\
(-0.5, 1) & ($n+1$, 1) & (0.0813, 0.0580, 0.4063, 0.8125) & -5.4321 & 1 \\
& & (0.1182, 0.0251, 0.1748,  0.0285)& -0.4781 & 0.3285\\
\dots & \dots & \dots & \dots & \dots \\
(0, 1) & ($n+1$, 0) & (0.0625, 0.0208, 0.2500, 0.2500)& -1.3333 & 1\\
(0, 0.2) & ($5n+1$, 0) & (0.1500, 0.0150, 0.2143, 0.0750)& -0.7779 & 1\\
 & & (0.00067, 0.0027, 0.2490, 0.0325) & 0.0735 & 0.0657\\
\dots & \dots & \dots & \dots & \dots \\
(0.5, 0.5) & ($n+1, n+1$)  & (0.0469, 0.0469, 0.1875, 0.1875) & -0.5625 & 1  \\
 &   & (0.0078, 0.0078, 0.1875, 0.0313) & 0.0625 & 0.2041 \\
 & & (0.0078, 0.0078, 0.0313, 0.1875) & 0.0625 & 0.2041 \\
\dots & \dots & \dots & \dots & \dots \\
(1,0 ) & ($n+1$, $n+1$) & (0.0625,  0.0625,  0.2500,  0.2500) & -1& 1 \\
 & & (0.0680, 0.0680, 0.1946, 0.1946) & -0.7608  & 0.9711\\
 \hline
\end{tabular}
\caption{Critical points for given parameters $a_i$ and Riemann surface genera $\frak{g}_i$ with $n \in \mathbb{Z}$. $\mathcal{K}$ denotes the Gaussian curvature of the Riemann surface and $\ell_2/\ell_1$ the degree to which the radius of the Riemann surface is squashed relative to the timelike direction. Expressions have been rounded to four decimal places, but can be found numerically to greater accuracy. Dots separate points corresponding to reductions on $H^2 \times T^2$, $H^2 \times S^2$ and $H^2 \times H^2$, respectively. The final entry corresponds to a point where supersymmetry of the $AdS_3$ vacuum is enhanced to $\mathcal{N} = (2,2)$.}
\end{center}
\end{table}

Our findings mirror the results for 3D gauged supergravities based on wrapped D3-brane geometries presented in \cite{Colgain:2015mta}. For parameters in the allowed ranges where good $AdS_3$ vacua exist, we either recover the unwarped $AdS_3$ vacuum, or in addition we find extra fixed-points. Solutions for sample points in parameter space are given in Table 1. For these new fixed-points, the geometry is supported by fluxes and the timelike direction is stretched, $\ell_1 > \ell_2$ leading to CTCs.  For all points in parameter space we have studied, we find that $\ell_1$ does not change, whereas $\ell_2$ decreases as we warp the geometry. Depending on the Gaussian curvature of the Riemann surface, the new solutions are either $\mathbb{R} \times H^2$ ($\mathcal{K} < 0$), which we know as G\"{o}del solutions \cite{Godel, Reboucas:1982hn}, or $\mathbb{R} \times S^2$ ($\mathcal{K} > 0$), which may be referred to as warped de Sitter \cite{Anninos:2009jt}.

To find these new warped solutions, we have solved (\ref{scalar_eqn}) for $W_I$, i. e. given ($a_1, a_2$) four equations for four unknowns. One can also attempt to find points where $\mathcal{K} =0$, where the spacetime would be topologically, $\mathbb{R}~ \times$ $T^2$, but in all the points we have studied, we have found no real solutions. This mirrors the wrapped D3-brane case \cite{Colgain:2015mta}, where such an outcome was shown not to arise.

However, in contrast to our expectations based on the analysis of ref. \cite{Colgain:2015mta}, for points in parameter space where the $AdS_3$ supersymmetry is enhanced to $\mathcal{N} = (2,2)$, for example ($a_1, a_2 ) \in \{ (0,1), (1,0) \}$, ($\frak{g}_1>0, \frak{g}_2>0)$, we find new fixed-points. This is hopefully evident from the last two entries of Table 1.

\section{Supergravity dual of $c$-extremization}
\label{sec:c_ext}
\setcounter{equation}{0}
In this section we review the results of $c$-extremization \cite{Benini:2012cz, Benini:2013cda}, a procedure to identify the exact R symmetry and central charge of a 2D theory with $\mathcal{N} = (0,2)$ supersymmetry. We recall that $\mathcal{N} = (0,2)$ SCFTs, like their $\mathcal{N}=1$ counterparts in 4D, possess a $U(1)$ R symmetry, which is associated with the right-movers in 2D. If there are additional Abelian flavour symmetries in the theory, it is well-known that there is an ambiguity in the R symmetry, since it is free to mix with other symmetries. The achievement of ref. \cite{Benini:2012cz, Benini:2013cda} is that the exact superconformal R symmetry is uniquely determined by extremising the trial $c$-function,
\be
\label{trial_c}
c_{R\, \textrm{trial}}(t_I) = 3 \left( k^{RR} + 2 \sum_{I} t_I k^{IR} + \sum_{IJ} t_I t_J k^{IJ} \right),
\ee
where $k^{RR}$ denotes the coefficient in the two-point function of the right-moving R current and $k^{IJ}$ are the coefficients of the flavour current two-point functions. Note that since $c_{R\, \textrm{trial}}$ is quadratic, it has a unique extremisation, leading to a procedure called $c$-extremization.

One setting where $c$-extremization plays a r\^{o}le is in the dimensional reduction of the 6D $\mathcal{N} = (2,0)$ theory associated to M5-branes on a product of Riemann surfaces, $\Sigma_{\frak{g}_1} \times \Sigma_{\frak{g}_2}$ to 2D, where it serves to identify the exact R symmetry and central charge. We here sketch the calculation, while referring the reader to ref. \cite{Benini:2013cda} for further details.

The 6D $\mathcal{N} = (2,0)$ theory has an $SO(5)$ R symmetry, so to preserve supersymmetry in the reduction, one twists the theory by turning on background gauge fields coupled to the $SO(2)^2$ Cartan subgroup of $SO(5)$. In the compactification, the trial R symmetry becomes a linear combination of the generators of $SO(2)^2$, $T_{A,B}$
\be
T_{R} = (1 + \epsilon) T_A + (1-\epsilon) T_B, \quad \epsilon \in \mathbb{R},
\ee
normalised so that a complex supercharge has R charge one. To determine the trial central charge (\ref{trial_c}), in the absence of a weakly-coupled Lagrangian formulation for 6D $\mathcal{N} = (2,0)$ theories, one can exploit the M5-brane anomaly polynomial \cite{Harvey:1998bx, Intriligator:2000eq, Yi:2001bz}
\be
I_{8} = \frac{r_{G}}{48} \left[ p_2(N)-p_2 (T) + \frac{1}{4} (p_1(T)-p_1(N))^2\right] + \frac{r_G h_G (h_G+1)}{24} p_2(N),
\ee
where $N$ and $T$ are the normal and tangent bundles, $p_i$ is the $i^{\textrm{th}}$ Pontryagin class \footnote{For a vector bundle $E$ over a differentiable manifold, $M$,
\bea
p_1 = \frac{1}{2} \left( \frac{i}{2 \pi} \right)^2 \textrm{tr} F^2, \quad p_2 = \frac{1}{8} \left( \frac{i}{2 \pi} \right)^4 (\textrm{tr} F^2 \wedge \textrm{tr} F^2 -2 \textrm{tr} F^4 ),
\eea
where $F$ denotes the curvature two-form, i. e. the background $SO(5)_{R}$ field strength.},
 and $r_G$ and $h_G$ are the rank and Coxeter number, respectively. For the $A_{N-1}$ theory, one has $r_G = N-1, d_G = N^2-1$ and $h_G = N$. Integrating the eight-form anomaly polynomial $I_{8}$ over $\Sigma_{\frak{g}_1} \times \Sigma_{\frak{g}_2}$, one compares with the anomaly polynomial of a 2D theory to determine $c_{R \, \textrm{trial}}$,
\be
I_{4} = \frac{c_{R}}{6} c_1 (F)^2 - \frac{c_R-c_L}{24} p_1 (T),
\ee
where $c_1$ denotes the first Chern class. Extremising $c_{R\,\textrm{trial}}$ to determine $\epsilon$, one plugs the expression back into $c_{R}$ to  determine the exact left and right central charges \cite{Benini:2013cda}
\bea
&&c_L = \frac{\eta_1 \eta_2}{4} \frac{d_G^2 h_G^2 \mathcal{P} +2 d_G h_G r_G (3 z_1^2 z_2^2 - 8 \kappa_1 \kappa_2 z_1 z_2 + \kappa_1^2 \kappa_2^2 )+ 3 r_G^2 z_1 z_2 (z_1 z_2 - 2 \kappa_1 \kappa_2) }{d_G h_G (\kappa_1 \kappa_2 -3 z_1 z_2) - 3 r_G z_1 z_2}, \nn
\label{c_exact} &&c_R = \frac{\eta_1 \eta_2}{4} \frac{d_G^2 h_G^2 \mathcal{P} +2 d_G h_G r_G (3 z_1^2 z_2^2 - 8 \kappa_1 \kappa_2 z_1 z_2 + \kappa_1^2 \kappa_2^2 )+ 9 r_G^2 \kappa_1 \kappa_2 z_1 z_2}{d_G h_G (\kappa_1 \kappa_2 -3 z_1 z_2) - 3 r_G z_1 z_2},
\eea
where we have defined
\be
\mathcal{P} = 3 z_1^2 z_2^2 + \kappa_1^2 z_2^2 + \kappa_2^2 z_1^2 - 8 \kappa_1 \kappa_2 z_1 z_2 + 3 \kappa_1^2 \kappa_2^2.
\ee
To make sense of the above expressions for $c_{L,R}$, we need to additionally define
\be
\eta_i = \biggl\{ \begin{array}{c} ~~~~1, ~~~~~~~\frak{g}_i = 1, \\ 2 | \frak{g}_i -1|, ~~\frak{g}_i \neq 1. \end{array}
\ee
The exact R symmetry is
\be
\label{exact_R}
T_R = T_A+T_B + \frac{d_G h_G (\kappa_1 z_2 + \kappa_2 z_1)}{d_G h_G (\kappa_1 \kappa_2-3 z_1 z_2)-3 r_G z_1 z_2} (T_A-T_B).
\ee

At the two-derivative level in supergravity, the goal is to recover the large $N$ limit of the exact central charge and R symmetry. To this degree of approximation, the central charge (\ref{c_exact}) and R symmetry (\ref{exact_R}) become:
\bea
\label{c_large_N}
&&c_{L} \simeq c_{R} \simeq 2 \eta_1 \eta_2 N^3 \frac{a_1^2 b_2^2+a_2^2 b_1^2 + a_1 a_2 b_1 b_2}{2 a_1 b_2 + 2 a_2 b_1-a_1 a_2 -b_1 b_2}, \\
\label{R_large_N}
&&T_{R} \simeq \frac{2 (a_1 b_2 + a_2 b_1- a_1 a_2)}{2 a_1 b_2 + 2 a_2 b_1 -a_1 a_2 -b_1 b_2} T_A+\frac{2 (a_1 b_2 + a_2 b_1- b_1 b_2)}{2 a_1 b_2 + 2 a_2 b_1 -a_1 a_2 -b_1 b_2} T_B,
\eea
where we have rewritten expressions using $z_i = a_i - b_i$ and the supersymmetry condition (\ref{susy_cond}) with $g_7 =2$. We remark that the subleading terms for the central charges, $c_{L}, c_{R}$, i.e. $\frac{1}{N^2}$ suppressed terms, have recently been matched at the four-derivative level \cite{Baggio:2014hua}.

Here we will show that the superpotential $T$ captures all information at the two-derivative level. We emphasise that this agreement does not stop at just the central charge and R symmetry, but the extremisation of $T$, which produces supersymmetric $AdS_3$ vacua, is the direct supergravity analogue of the trial central charge in field theory that one extremises to find the exact result.

 It is now timely to recast our supergravity action in terms of the canonical expressions for a 3D $\mathcal{N} =2$ gauged supergravity \cite{deWit:2003ja}. To do so, we recall some salient details;  firstly, the superpotential is quadratic in the moment maps $\mathcal{V}^I$ associated to gauged isometries
\be
T = 2 \mathcal{V}^I \Theta_{IJ} \mathcal{V}^J,
\ee
which is further given in terms of the embedding tensor $\Theta_{IJ}$. Secondly, it is the embedding tensor that determines the CS term \footnote{Here we just focus on the Abelian case.}:
\be
\mathcal{L}_{\textrm{CS}} = \frac{1}{2} \mathcal{A}^I \Theta_{IJ} \dd \mathcal{A}^J.
\ee
Writing the gauge fields, $\mathcal{A}^I$, $I = 1,\dots, 4$ in the order $B^1, B^2, A^1$ and $A^2$ respectively, thus making connection with the gauge fields of section \ref{sec:KKred}, we can read off the components of the embedding tensor
\bea
\Theta_{12} &=& - \frac{1}{4g_7} (a_1 b_2 + a_2 b_1), ~~\Theta_{13} = - \frac{1}{2} b_2, ~~\Theta_{14} = -\frac{1}{2} a_2, \nn
\Theta_{34} &=& g_7, \quad \quad~~~~~~~~~~~~~~~~~\Theta_{23} = - \frac{1}{2} b_1, ~~ \Theta_{24} = - \frac{1}{2} a_1.
\eea
This then determines $T$ once we identify the associated moment maps:
\be
\mathcal{V}^I = \frac{1}{4} e^{-W_I}
\ee
To fully specify $T$, we should introduce $\mathcal{V}^0 = 1$, which is associated to a central extension of the isometry group that generates the $SO(2)$ R symmetry. The additional components of the embedding tensor are $\Theta_{I0} = \frac{g_7}{2}$.

It should be clear from the above analysis that the inverse of the superpotential $T$ may be regarded as the trial $c$-function in the vicinity of the superconformal fixed-point. Not only is it extremised at the $AdS_3$ vacuum, but it is also quadratic in moment maps, mirroring the trial $c$-function (\ref{trial_c}). Moreover, as we have just seen, since $\ell \sim \frac{1}{T}$, we have from the Brown-Henneaux formula \cite{Brown:1986nw}, $ c=  \frac{3}{2} \frac{\ell}{G_3}$, that the central charge $c$ is inversely proportional to $T$. In fact, it is known that the r\^{o}le of $T$ is more general; it is the natural (super)gravity analogue of the Zamolodchikov $c$-function \cite{Zamolodchikov:1986gt} for holographic RG flows interpolating between $AdS_3$ vacua \cite{Berg:2001ty}. It has recently been noted \cite{Colgain:2015ela} that it also decreases in flows from $AdS_3$ to G\"{o}del \cite{Godel} fixed-points.

To make comparison, we need to fix the constant of proportionality, an exercise that is most easily performed by borrowing the conventions of
of \cite{Maldacena:2000mw} and comparing to the Brown-Henneaux formula. Doing so, we recover the result of \cite{Benini:2013cda} \footnote{In the conventions of \cite{Maldacena:2000mw}, the 11D Newton's constant is $G^{11}_{N} = 16 \pi^7 \ell_p^9$, the AdS$_7$ radius is taken to be one, $R_{AdS_7} = 2 (\pi N)^{\frac{1}{3}} \ell_p =1$. As a direct consequence of the choice of radius, the gauge coupling of the SO(5) gauged supergravity becomes $m=2$ and $G^7_{N} =  (3 \pi^2/16) N^{-3}$. },
\be
\label{sugra_c}
c = \frac{1}{2} ( c_L + c_R ) \simeq \frac{3 \ell}{~2 G^3_{N}} \simeq \frac{16 \eta_1 \eta_2}{T} N^3,
\ee
which agrees with (\ref{c_large_N}) when $g_7=2$. Note, this result holds for the $A_{N}$ theory. To get the result for the $D_{N}$ theory, we can simply consider the $\mathbb{Z}_2$ orbifold of flat spacetime with M5-branes  at the origin. As a result, we have $c_{L}^{D_N} \simeq c_{R}^{D_N} \simeq 4 c_{R}^{A_N}$.

We can also extract the R symmetry from 3D supergravity in the large $N$ limit. To do so, we recall that the R symmetry evaluated at the $AdS_3$ fixed-point is given by the linear combination \cite{Karndumri:2013iqa} \footnote{See \cite{Tachikawa:2005tq} for the 5D analogue.}
\be
\label{sugra_R}
R =  \frac{2 \mathcal{V}^{I}}{T} Q_{I},
\ee
where $Q_{I}$ denotes the charges corresponding to the $U(1)$ currents. From the higher-dimensional supergravity perspective, the R symmetry is a linear combination of the two gauged $U(1)$ isometries with gauge fields $B^I$, i. e. of the gauged $SO(2)^2$ Cartan of the maximally supersymmetric 7D theory. As can be seen from (\ref{cov_diff1}) and (\ref{cov_diff2}), these gauge fields are dualised into $U(1)$ isometries in our scalar manifold, $Y_I$, which enjoy a shift symmetry $Y_I \rightarrow Y_I+ \alpha_I$, where $\alpha_I$ denote constants. We can thus extract the R symmetry from the moment maps $\mathcal{V}^I = \frac{1}{4} e^{-W_I}$, $I = 1, 2$ associated to the gauging of these isometries. The result is
\bea
\frac{2 \mathcal{V}_1 }{T} =  \frac{2 (a_1 b_2 + a_2 b_1 -a_1 a_2)}{(2 a_1 b_2 +2 a_2 b_1-a_1 a_2-b_1 b_2)g_7}, ~~ \frac{2 \mathcal{V}_2 }{T} = \frac{2 (a_1 b_2 + a_2 b_1 -b_1 b_2)}{(2 a_1 b_2 +2 a_2 b_1-a_1 a_2-b_1 b_2)g_7}.
\eea
We note that this agrees perfectly with (\ref{R_large_N}) when $g_7=2$ and one takes into account the factor of one-half in (\ref{fluxes}) between $F^{I}$ and $G^{I}$, since the moment map associated to $F^I$ is $\frac{1}{2} e^{-W_I}$.

As mentioned earlier, by tuning the parameters $a_1 = a_2 = a$, $b_1 = b_2 =b$, and consequently, $\kappa_1 = \kappa_2 = \kappa$, it is now easy to determine the corresponding central charge and R symmetry for M5-branes wrapped on a KE$_4$ manifold. With $g_7=2$, we immediately see from extremising $T$ that $\kappa < 0$, so that the four-manifold is negatively curved. Once again, we normalise so that $\kappa =-1$.

The central charge and R symmetry then follow from (\ref{sugra_c}) and (\ref{sugra_R})
\bea
c &\simeq& \frac{a^2 b^2}{4 a b - a^2 -b^2} \frac{3 \vol(\textrm{KE}_4) N^3}{2 \pi^2}, \nn
T_{R} &\simeq& \frac{2 (2 b -a)}{4 a b - a^2 -b^2}T_A +  \frac{2 (2 a -b)}{4 a b - a^2 -b^2}T_B,
\eea
and are simply a refinement of previous expressions, once one takes account of the fact that $\eta_1 \eta_2 = \frac{1}{4 \pi^2}\vol(\Sigma_1 \times \Sigma_2) = \frac{1}{4 \pi^2}\vol(\textrm{KE}_4) $ and $a+b =1$. As expected, both of these agree with the exact central charge and R symmetry in the large $N$ limit \cite{Benini:2013cda}.

\section{11D uplift of $AdS_3$ vacua}
\label{sec:11Duplift}
\setcounter{equation}{0}
As demonstrated in section \ref{sec:KKred}, any solution to our 3D $\mathcal{N} = 2$ gauged supergravity may be viewed as a solution to 11D supergravity \cite{Nastase:1999cb, Nastase:1999kf}. In this section, we focus on the uplifts of supersymmetric $AdS_3$ vacua corresponding to the extrema of the superpotential, which we will write in canonical form as a $U(1)$ fibration over a 6D $SU(3)$-structure manifold  \cite{Figueras:2007cn, Gauntlett:2006ux, MacConamhna:2006nb}.  

To perform the uplift, it is easiest to make use of the results of ref. \cite{Cvetic:2000ah}, which are already tailored to the $U(1)^2$ truncation of the 7D theory. In the process, we adopt the following parametrisation for the (constrained) $S^4$ scalars,
\be
\mu_5 = \rho_0, \quad \mu_{2 \alpha-1} = \rho_{\alpha} \sin \phi_{\alpha}, \quad \mu_{2 \alpha} = \rho_{\alpha} \cos \phi_{\alpha},
\ee
where $\alpha = 1, 2$ and $\phi_{\alpha}$ are $2 \pi$-periodic.  We observe that since the $\mu_i$ are constrained so that $\sum_{i=1}^5 \mu_i^2 =1$, we necessarily have $\sum_{\alpha}^2 \rho_{\alpha}^2 =1$, so that $\rho_{\alpha} = 0, 1, 2$ now parametrise an $S^2$. More concretely, we can choose,
\be
\rho_0 = \cos \alpha, \quad \rho_1 = \sin \alpha \cos \beta, \quad \rho_2 = \sin \alpha \sin \beta.
\ee
In terms of the coordinates $\alpha, \beta$ parametrising the $S^2$, the uplifted 11D metric for $AdS_3$ vacua may be expressed as
\bea
\label{11D_uplift}
\dd s^2_{11} &=& \Delta^{\frac{1}{3}} \left[ e^{-4(\lambda_3 + \lambda_4)} \dd s^2(AdS_3) + e^{2 \lambda_3} \dd s^2(\Sigma_{\frak{g}_1}) + e^{2 \lambda_4 } \dd s^2(\Sigma_{\frak{g}_2}) \right] \\ &+& \frac{1}{4} \Delta^{-\frac{2}{3}} \biggl[ e^{4 \l_1 +4  \l_2} \frac{\Delta}{X} \dd \alpha^2 + e^{-2(\lambda_1 + \lambda_2)} X \DD \beta^2 + \sin^2 \alpha \left( e^{-2 \lambda_1} \cos^2 \beta \DD \phi_1^2 + e^{-2 \lambda_2} \sin^2 \beta \DD \phi_2^2 \right)  \biggr], \nonumber
\eea
where $g_7 = 2$, the warp factor is now $\Delta =   e^{-4 \lambda_1 -4 \lambda_2} \cos^2 \alpha+ \sin^2 \alpha X$, and we have further defined
\bea
\DD \phi_{\alpha} &=& \dd \phi_{\alpha} + 4 A^{\alpha}, ~~\alpha = 1, 2, \nn
X &=&  e^{2 \lambda_1} \cos^2 \beta + e^{2 \lambda_2} \sin^2 \beta, \nn
\DD \beta &=& \sin \alpha \dd \beta + \frac{(e^{2 \lambda_1} - e^{2 \lambda_2})}{X} \cos \alpha \cos \beta \sin \beta \dd \alpha.
\eea

The four-form flux may be expressed as
\bea
\label{G4}
G_4 &=& \frac{1}{8} U \Delta^{-2} \sin^2 \alpha \cos \beta \sin \beta \DD \phi_1 \DD \phi_2 \dd \alpha \DD \beta \nn
&+& \frac{\Delta^{-1}}{2} F^2 \DD \phi_1 \left[ \sin \alpha \cos^2 \beta e^{2 \lambda_1} \frac{\Delta}{X} \dd \alpha - e^{-4 \lambda_1 - 4 \lambda_2} \cos \alpha \sin \alpha \cos \beta \sin \beta \DD \beta \right] \nn
&+&  \frac{\Delta^{-1}}{2} F^1 \DD \phi_2 \left[ \sin \alpha \sin^2 \beta e^{2 \lambda_2} \frac{\Delta}{X} \dd \alpha + e^{-4 \lambda_1 - 4 \lambda_2} \cos \alpha \sin \alpha \cos \beta \sin \beta \DD \beta \right] \nn
&+& \frac{1}{16} ( a_1 b_2 + a_2 b_1) \left[ 2 \cos \alpha  \vol(\Sigma_{\frak{g}_1}) \vol(\Sigma_{\frak{g}_2}) - \sin \alpha e^{4(\lambda_1 + \lambda_2) -8 (\lambda_3 + \lambda_4)} \vol (AdS_3) \dd \alpha  \right],
\eea
where we have omitted obvious wedge products to save space and defined
\bea
U &=& \cos^2 \alpha \left( e^{-8 \lambda_1 - 8 \lambda_2}-2 e^{-2 \lambda_1-4 \lambda_2} -2 e^{-2 \lambda_2 -4 \lambda_1} \right) \nn
&-& \sin^2 \alpha \cos^2 \beta \left( 2 e^{2 \lambda_1 + 2 \lambda_2} + e^{-2 \lambda_1 -4 \lambda_2} \right) - \sin^2 \alpha \sin^2 \beta \left( 2 e^{2 \lambda_1 + 2 \lambda_2} + e^{-4 \lambda_1 - 2 \lambda_2}\right).
\eea

Imposing the Bianchi identity $\dd G_4 = 0$ leads, in the notation of the earlier section, to the constraints:
\be
\dd C = \frac{4}{g} F^1 \wedge F^2  + 2 \sqrt{3} g \, e^{-4 \lambda_1 -4 \lambda_2} *_7 C = 0.
\ee
One observes that these constraints are indeed satisfied for $AdS_3$ vacua, when the expressions for the field strengths $F^I$ and three-form potential (\ref{fluxes}), (\ref{rho}) are inserted \footnote{The contribution to $G_4$ due to $C$ was omitted in \cite{Benini:2013cda}, so the Bianchi identity will not be satisfied.}.

Having uplifted the geometry, we can now comment on how the twist parameters $a_i, b_i$ should be quantised, so that the geometry is well-defined. Demanding that the gauge field is a connection on a bona fide $U(1)$ fibration, we require that the periods of the first Chern class be integer valued. This leads to the conditions
\be
\label{quant}
\frac{1}{2 \pi} \int_{\Sigma_{\frak{g}}} 2  g_7 \, \dd A^{1} = g_7 \, a_{i} (\frak{g}-1) \in \mathbb{Z}, \quad \frac{1}{2 \pi} \int_{\Sigma_{\frak{g}}} 2  g_7 \, \dd A^{2} = g_7 \, b_{i} (\frak{g}-1) \in \mathbb{Z},
\ee
where $\frak{g} \neq 1$ is the genus of the Riemann surface, $\Sigma_{\frak{g}}$, over which we integrate. When one compactifies on a torus, $\frak{g}=1$, this condition simply reads $a_i, b_i \in \mathbb{Z}$. As such, we recognise the need to quantise the parameters so that the internal geometry is well-defined. This requirement places stringent constraints on reductions on spheres and tori, however in the case of compactifications on hyperbolic spaces, one is free to quotient the Riemann surface without breaking supersymmetry, thus increasing the genus. 

We can now use the result of the previous section, namely (\ref{sugra_R}), to rewrite (\ref{11D_uplift}) in the canonical form in order to distinguish the R symmetry $U(1)_R$ from the global symmetry $U(1)_G$. From (\ref{sugra_R}), we see that the R symmetry vector, $K$, is
\be
K = \frac{e^{-W_1}}{T} \partial_{\phi_1} + \frac{e^{-W_2}}{T} \partial_{\phi_2} = 2 \ell e^{-2 (\lambda_3 + \lambda_4)} \left( e^{2 \lambda_1} \partial_{\phi_1} + e^{2 \lambda_2} \partial_{\phi_2}\right)
\ee
where $\partial_{\phi_i}$ are vectors associated to the $U(1)$ isometries. Furthermore, we have rewritten $T$ in terms of the $AdS_3$ radius, $T = 1/(2 \ell)$ and $W_I$ in terms of $\lambda_I$ (\ref{scalar_redef}). Using the 11D metric (\ref{11D_uplift}), one can determine the dual one-form
\be
K = \frac{\ell \sin^2 \alpha}{2 e^{2 (\lambda_3 + \lambda_4)} \Delta^{\frac{2}{3}}} \left( \cos^2 \beta \DD \phi_1 + \sin^2 \beta \DD \phi_2 \right).
\ee
Normalising $K$ to unit norm, we can rewrite the metric in the $\phi_i$-directions in the following fashion:
\bea
&&\dd s^2_2 = \frac{\sin^2 \alpha}{4\,  \Delta^{\frac{2}{3}}} \left( e^{-2 \lambda_1}  \cos^2 \beta \,  \DD \phi_1^2+ e^{-2 \lambda_2} \sin^2 \beta \, \DD \phi_2^2 \right) \nn
\label{metric1}
&&=\frac{\sin^2 \alpha}{4 \, \Delta^{\frac{2}{3}} \, X} \left[ \left( \cos^2 \beta \DD \phi_1 + \sin^2 \beta \DD \phi_2 \right)^2 + \cos^2 \beta \sin^2 \beta \left( e^{\lambda_2 - \lambda_1} \DD \phi_1 - e^{\lambda_1 - \lambda_2} \DD \phi_2 \right)^2 \right].
\eea
We note that the first term on the RHS is simply the one-form corresponding to the R symmetry, while the remaining term is the $U(1)_G$ symmetry. 

We are free to change coordinates as follows:
\be
\dd \phi_1 = e^{2 \lambda_1} ( \dd \psi_{1} + \dd \psi_2), \quad \dd \phi_2 = e^{2 \lambda_2} ( \dd \psi_1 - \dd \psi_2),
\ee
where we observe for $AdS_3$ solutions that $\lambda_I$ are simply constant. As a result of this redefinition, the 11D  metric (\ref{11D_uplift}) becomes
\bea
\dd s^2_{11} &=& \Delta^{\frac{1}{3}} \left[ e^{-4(\lambda_3 + \lambda_4)} \dd s^2(AdS_3) + e^{2 \lambda_3} \dd s^2(\Sigma_{\frak{g}_1}) + e^{2 \lambda_4 } \dd s^2(\Sigma_{\frak{g}_2}) \right] \nn &+& \frac{1}{4} \Delta^{-\frac{2}{3}} \biggl( e^{4 (\l_1 + \l_2)} \, \frac{\Delta}{X} \, \dd \alpha^2 + e^{-2(\lambda_1 + \lambda_2)} \, X \, \DD \beta^2 \\ &+& \sin^2 \alpha \biggl[ X \left( \DD \psi_1 + \frac{(e^{2 \lambda_1} \cos^2 \beta - e^{2 \lambda_2} \sin^2 \beta)}{X} \DD \psi_2  \right)^2 +  \frac{ e^{2(\lambda_1 + \lambda_2)}}{X} \sin^2( 2 \beta) \DD \psi_2^2 \biggr]  \biggr), \nonumber
\eea
which is the canonical form for $U(1)_R \times U(1)_G$, where $\partial_{\psi_1}$ is the R symmetry vector and $\partial_{\psi_2}$ the global symmetry vector and we have defined
\be
\DD \psi_1 = \dd \psi_1 + 2 e^{-2 \lambda_1} A^1 + 2 e^{-2 \lambda_2} A^2, \quad \DD \psi_2 = \dd \psi_2 +2 e^{-2 \lambda_1} A^1 - 2 e^{-2 \lambda_2} A^2.
\ee
It is straightforward to recast the four-form flux, $G_4$, in the new coordinates $\psi_i$ and we omit the expression.

At this stage we could contemplate performing a dimensional reduction on the $U(1)_G$ isometry, however this would lead to a singular geometry where the dilaton $\Phi$ blows at $\alpha = 0$. One could further choose one of the Riemann surfaces to be a torus and perform a T-duality without breaking supersymmetry, but the singularity will persist \footnote{In the the process of twisting and compactifying the Killing spinor become independent of the Riemann surfaces. One can then infer from ref. \cite{Kelekci:2014ima} that supersymmetry will be preserved in the T-duality.}, so it is better to consider the 11D geometries, which are regular.

As a consistency check on the uplifted geometry (\ref{11D_uplift}) and the identification of the R symmetry in 11D, we should compare with existing classifications. To the extent of our knowledge, the most general existing classifications of supersymmetric $AdS_3$ solutions of 11D supergravity, with $\mathcal{N} = (0,2)$ dual SCFTs, can be found in ref. \cite{Gauntlett:2006ux} and ref. \cite{Figueras:2007cn}. Since \cite{Gauntlett:2006ux} considers purely magnetic flux configurations, we focus on the \cite{Figueras:2007cn}. It has already been checked that a number of known solutions to 7D supergravity \cite{Gauntlett:2000ng,Gauntlett:2001jj} fall into these classifications. Supersymmetric geometries in this general class take the form \cite{Figueras:2007cn}
\bea
\label{SUSY_geom}
\dd s^2_{11} &=& \frac{1}{\lambda m^2} \left[  \dd s^2(AdS_3) + \frac{\lambda^3}{4 \sin^2 \theta} \dd \rho \otimes \dd \rho \right] + e^7 \otimes e^7 + \dd s^2 (\mathcal{N}_6), \nn
F_4 &=& \frac{1}{m^3} \vol_{AdS_3} \wedge \dd [ \rho - \lambda^{-3/2} \cos \theta]  \nn
&+& \frac{\lambda^{3/2}}{\sin^2 \theta} ( \cos \theta + *_8) (\dd [ \lambda^{-3/2} \sin \theta J \wedge e^7] - 2 m \lambda^{-1} J \wedge J) + 2 m \lambda^{1/2} J \wedge e^{7} \wedge \hat{\rho},
\eea
where $\mathcal{N}_6$ admits an $SU(3)$ structure, which along with $\lambda$, $\theta$,  is independent of $AdS_3$ coordinates and we have defined $\hat{\rho} = \lambda/(2 m \sin \theta) \dd \rho$. In terms of the $SU(3)$ structure forms, $J$ and $\Omega$, the supersymmetry conditions may be expressed as \cite{Figueras:2007cn} \footnote{Here $A \lrcorner B = \frac{1}{p!} A_{\mu_1 \dots \mu_p} B^{\mu_1 \dots \mu_p}$ for $p$-forms $A$ and $B$.},
\bea
\label{susy_cond_class}
\hat{\rho} \wedge \dd (\lambda^{-1} J \wedge J)  &=& 0, \\
J \lrcorner \dd e^7 &=& \frac{2 m \lambda^{\frac{1}{2}}}{\sin \theta} (2 - \sin^2 \theta) - \cos \theta \hat{\rho} \lrcorner \dd \log \left( \frac{\lambda^{\frac{3}{2}} \cos \theta}{ \sin^2 \theta} \right), \\
\dd ( \lambda^{-\frac{3}{2}} \sin \theta \im \Omega) &=& 2 m \lambda^{-1} ( e^7 \wedge \re \Omega - \cos \theta \hat{\rho} \wedge \im \Omega). 
\eea
The combination $ \lambda^{-3/2} \cos \theta$ just depends on the coordinate $\rho$, $f (\rho) =  \lambda^{-3/2} \cos \theta$, and setting $f(\rho) = \rho$, we recover purely magnetic solutions and the results of ref. \cite{Gauntlett:2006ux}. 

To make comparison, we note that the metric and electric flux take the same form provided 
\be
\rho = \ell^{-1} e^{2 (\lambda_1 + \lambda_2) - 4 (\lambda_3 + \lambda_4)} \cos \alpha, \quad f = e^{-2 (\lambda_1 + \lambda_2)-6(\lambda_3 + \lambda_4)} \cos \alpha. 
\ee
Furthermore, we have checked that 
\bea
e^{7} &=& \frac{\sin \alpha}{2 \Delta^{\frac{1}{3}} \sqrt{X}} (\cos^2 \beta \DD \phi_1 + \sin^2 \beta \DD \phi_2), \nn
J &=& \Delta^{\frac{1}{3}} \left[ e^{2 \lambda_3} \vol(\Sigma_{\frak{g}_1}) +  e^{2 \lambda_4} \vol(\Sigma_{\frak{g}_2}) \right] + \frac{e^{-(\lambda_1 + \lambda_2)} \sin \alpha \sin 2\beta}{8 \Delta^{\frac{2}{3}} } \left( e^{\lambda_2 - \lambda_1} \DD \phi_1 - e^{\lambda_1 - \lambda_2} \DD \phi_2 \right) \wedge \DD \beta, \nn
\Omega &=& \frac{1}{2} e^{- \lambda_1 - \lambda_2 + \lambda_3 + \lambda_4} (\bar{e}^1 + i \bar{e}^2) \wedge (\bar{e}^3 + i \bar{e}^4) \wedge \left[ \frac{ \sin \alpha \sin 2 \beta}{2 \sqrt{X}} (e^{2 \lambda_2} \DD \phi_1 - e^{2 \lambda_1} \DD \phi_2) + i \sqrt{X} \DD \beta \right], \nn
\eea
where $\bar{e}^i$ denote appropriately chosen one-forms on the Riemann surfaces, satisfy the above supersymmetry conditions through a number of non-trivial cancellations. Just highlighting one particular case, we observe that the simplest condition (\ref{susy_cond_class}) implies the following constraint on the scalars
\be
e^{W_1} (a_1 e^{W_3} + a_2 e^{W_4}) = e^{W_2} (b_1 e^{W_3}+b_2 e^{W_4}) + 8 e^{W_3 + W_4} (e^{W_1}-e^{W_2}). 
\ee
We emphasise that it is not immediately obvious that this condition is satisfied, since this equation is an artifact of the 11D description. However, it can be checked using (\ref{scalar_redef}) that this condition is satisfied for all $a_i, b_i$. This agreement provides yet another consistency check on the results of ref. \cite{Figueras:2007cn} and also allows us to confirm that the R symmetry from the 3D perspective agrees with the canonical R symmetry from 11D supergravity, in line with our expectations. 

Finding that the supersymmetry conditions are satisfied for all twistings $a_i, b_i$ is not entirely unexpected. The reason being that the classification of ref. \cite{Figueras:2007cn}, although it assumes a wrapped M5-brane ansatz from the outset, recovers a known classification of all minimally supersymmetric $AdS_3$ solutions in 11D supergravity \cite{Martelli:2003ki}. Since classifications with different supersymmetry are simply related via an identification of $G$ structures - in this case two orthogonal $G_2$-structures define the $SU(3)$-structure - it may be expected that (\ref{SUSY_geom}) is sufficiently general to cover the uplifted 11D geometries. \footnote{Recently,  $AdS_3$ solutions with $\mathcal{N} = (0,2)$ supersymmetry \cite{Araujo:2015npa, Bea:2015fja} have been generated via $SU(2)$ non-Abelian T-duality \cite{Itsios:2013wd} (also \cite{Kelekci:2014ima}). It would also be interesting to confirm that these fit into the above classification. It was reported recently that another non-Abelian T-dual $AdS_3$ geometry with $\mathcal{N} = (0,4)$ supersymmetry \cite{Lozano:2015bra} fell outside of one of these classes.} 

Finally, one last comment. While we have focused on 11D uplifts in this section, recently it has been shown \cite{Passias:2015gya} how minimal gauged supergravity \cite{Townsend:1983kk} in 7D may be embedded in massive IIA supergravity. Since minimal gauged supergravity has $SU(2)$ valued gauge fields, this necessitates that one truncates the $U(1)^2$ theory to $U(1)$, leaving a single gauge field and scalar in 7D. As a result, only solutions based on K\"ahler-Einstein compactifications from section \ref{sec:KKred},  where $a_1 = a_2$, $b_1 = b_2$, can be embedded this way.

\section{Discussion}
If there is a take-home message from our work, it should be the observation that, for the class of 3D gauged supergravities arising from wrapped M5-branes, the superpotential $T$ contains a wealth of information. Extremising it, we get the $AdS_3$ vacua, and the Brown-Henneaux central charge is, modulo a coefficient, simply the inverse of $T$. Indeed, the extremisation process through which one arrives at $AdS_3$ vacua deftly encapsulates the $c$-extremization  procedure from 2D CFTs \cite{Benini:2012cz,Benini:2013cda}. To add weight to this statement, one should recall that the R symmetry in the large $N$ limit is given in terms of the moment maps, which appear in $T$ quadratically contracted with the CS coefficients (embedding tensor). We have checked that the R symmetry agrees with the canonical $U(1)$ R symmetry from known classifications. Moreover, as we have seen in section \ref{sec:timelike}, $T$ dictates all the supersymmetric solutions in the theory and as expressions are presented implicitly in terms of $T$, the analysis should hold for all timelike supersymmetric solutions to 3D $\mathcal{N}=2$ gauged supergravity, once the explicit superpotential is determined.  

Although we have not touched upon it in this work, $T$ also gives a concrete prediction for the central charge at the warped $AdS_3$ fixed-point. This claim is based on the assumption that we can use the same coefficient that reproduces the Brown-Henneaux result for $AdS_3$ vacua. It would be interesting to see if the same result may be recovered from the asymptotic symmetry algebra by generalising the analysis of ref. \cite{Compere:2007in} to theories with scalar potentials. 

Unfortunately, all comments above are restricted to the two-derivative level and it would be interesting if one could find a higher-derivative analogue of the superpotential in 3D that also encodes the corrections. From a 5D perspective \cite{Baggio:2014hua}, we already know the corrected $AdS_3$ supersymmetric geometries, and that the CFT result may be recovered at subleading order using $c$-extremization of Kraus-Larsen \cite{Kraus:2005vz}, while the difference between $c_L$ and $c_R$, which is not evident at the two-derivative level, can be read off from the gravitational CS terms in 3D \cite{Kraus:2005zm}. It would be nice to streamline this process using supersymmetry, if possible. By either dimensional reduction of 5D off-shell supergravity, or working directly with known supersymmetric invariants in 3D, for example \cite{Alkac:2014hwa}, one should be able to identify the analogue of $T$ with four-derivatives in order to see to what extent corrected solutions may be found via an extremisation process. As a warm-up, it should be interesting to match the R symmetry at subleading order to reconcile the prescription in the literature \cite{Hanaki:2006pj} with the analysis of ref.  \cite{Baggio:2014hua}, which already agrees with the expected CFT central charges. 

It is a common feature that the 3D gauged supergravities, which we have found via dimensional reduction, all have non-compact target spaces. In principle, 3D $\mathcal{N} =2$ gauged supergravity also allows for compact target spaces, such as $\mathbb{C} \textrm{P}^1$ \cite{Deger:1999st} and more generally $\mathbb{C} \textrm{P}^n$ \cite{Gukov:2015qea}.  Identifying an embedding for these theories would help elucidate properties of the dual $\mathcal{N} = (0,2)$ SCFT and allow one to study RG flows from both the perspective of field theory and supergravity.  We hope to explore this in future work. 

\section*{Acknowledgements}
We are grateful to J. Gutowski, H. Lin, D. Robbins, E. Sezgin, W. Sabra, A. Tomasiello \& H. Yavartanoo for discussion. P. K. is supported by Chulalongkorn University through
Ratchadapisek Sompoch Endowment Fund under grant Sci-Super 2014-001. E. \'O C is supported by the Marie Curie award PIOF-2012-328625 T-DUALITIES. E \'O C is grateful to Matheson (Dublin) for providing stationary. 
\appendix

\section{Reduction of supersymmetry variations}
In this appendix, we reduce the supersymmetry variations of the $U(1)^2$ truncation of 7D $SO(5)$ gauged supergravity on a product of Riemann surfaces. A similar reduction on a single Riemann surface from 5D $U(1)^3$ gauged supergravity appeared in \cite{Colgain:2015mta}. In 7D the supersymmetry variations read \cite{Pernici:1984xx}:
\bea
\delta \psi_{\mu} &=& \biggl[ \nabla_{\mu} + \frac{1}{4} Q_{\mu \, ij} \Gamma^{ij} + \frac{g_7}{20} T \gamma_{\mu} - \frac{1}{40} (\gamma_{\mu}^{~\nu \rho}-8 \delta_{\mu}^{\nu} \gamma^{\rho}) \Gamma_{ij} \Pi^i_{A} \Pi^j_{B} F^{AB}_{\nu \rho} \nn
&& \phantom{xxxxxxxxxx} + \frac{g_7}{10 \sqrt{3}} (\gamma_{\mu}^{~\nu \rho \sigma} - \frac{9}{2} \delta^{\nu}_{\mu} \gamma^{\rho \sigma}) \Gamma^i (\Pi^{-1})^{A}_i S_{A \, \nu \rho \sigma} \biggr] \epsilon, \nn
\delta \chi_i &=& \biggl[ \frac{1}{2} P_{\mu i j} \gamma^{\mu} \G^{j} + \frac{g_7}{2} ( T_{ij} - \frac{1}{5} \delta_{ij} T) \Gamma^j  + \frac{1}{16} (\Gamma_{kl} \Gamma_i - \frac{1}{5} \Gamma_i \Gamma_{kl} ) \gamma^{\mu \nu} \Pi^{k}_{A} \Pi^{l}_{B} F_{\mu \nu}^{AB} \nn
&& \phantom{xxxxxxxxxx} + \frac{g_7}{20 \sqrt{3}} \gamma^{\mu \nu \rho} ( \Gamma_{i}^{~j} - 4 \delta_{i}^{~j} )  (\Pi^{-1})^{A}_j  S_{A \, \mu \nu \rho} \biggr] \epsilon,
\eea
where $\epsilon$ denotes the 7D supersymmetry variation, $T = \delta^{ij} T_{ij}$, and we have further defined
\bea
\Pi^{i}_{A} &=& \textrm{diag} ( e^{-\lambda_1}, e^{-\lambda_1}, e^{-\lambda_2}, e^{-\lambda_2}, e^{2 \lambda_1 + 2 \lambda_2}), \nn
M_{\mu i j} &=& (\Pi^{-1})^{A}_{~i} ( \delta_{A}^{~B} \partial_{\mu} + 2 g_7 A_{\mu\, A}^{~~~B} ) \Pi^{k}_{~B} \delta_{kj}, \nn
Q_{\mu ij} &=& \frac{1}{2} ( M_{\mu \, ij} - M_{\mu \, ji} ), \quad P_{\mu ij} = \frac{1}{2} (M_{\mu \, ij} + M_{\mu \, ji}).
\eea
The corresponding expression for $T_{ij}$ appears in the text (\ref{Tij}).

To perform the reduction, we decompose the 7D supersymmetry variation and gamma matrices as
\bea
\epsilon &=& e^{C} \xi \otimes \eta_1 \otimes \eta_2, \nn
\gamma^{a} &=& \rho^{a} \otimes \sigma^3 \otimes \sigma^3, \nn
\gamma^{m+2} &=& 1 \otimes \sigma^{m} \otimes \sigma^3, \nn
\gamma^{m+4} &=& 1 \otimes 1 \otimes \sigma^{m},
\eea
where now $a = 0, 1, 2$, $m = 1, 2$ and $C$ is a 3D scalar, yet to be determined. We impose the projection conditions
\be
\label{red_projection}
\g_{34}\epsilon=  \gamma_{56} \epsilon = -\G^{12} \e = -\G^{34} \e  = i \e,
\ee
where $\g_i$ are tangent space gamma matrices and $\G^i$ denote $SO(5)$ gamma matrices. Note we also take $\gamma_{0123456} = 1$ and $\G^{12345} = 1$, so this implies $ \gamma^{012} \e = \e$. With four projection conditions, we are generically left with two supersymmetries.

Performing the reduction using the ansatz in the text,  one finds the following algebraic conditions:
\bea
2 \G^1 \delta \chi_1 &=& e^{2(\l_3+\l_4)} \biggl[ -    \frac{1}{10} {\slashed \partial } (-3 W_1 +2 W_2 + W_3 + W_4) + \frac{g_7}{5} \left( 3 e^{-W_1} - 2 e^{-W_2} - e^{-W_3-W_4} \right)  \nn &+&
\frac{1}{20} \left( 3 a_1 e^{-W_2-W_4} + 3 a_2 e^{-W_2 - W_3} - 2 b_1 e^{-W_1-W_4} - 2 b_2 e^{-W_1-W_3} \right), \nn &-& \frac{1}{40 g_7} (a_1 b_2 + a_2 b_1) e^{-W_1-W_2}- i\frac{3 g_7}{10 \sqrt{3}}(e^{-W_4} \slashed{c}_1 + e^{-W_3} \slashed{c}_2)  \nn &+& i\frac{3}{20} e^{W_1} \slashed{G}^1 - i \frac{1}{10} e^{W_2} \slashed{G}^2  \biggr] \e , \nn
2 \G^3 \delta \chi_3 &=& e^{2 (\lambda_3 + \lambda_4)}\biggl[ -\frac{1}{10} {\slashed \partial (2 W_1 -3 W_2 + W_3 + W_4)}  + \frac{g_7}{5} \left( 3 e^{-W_2} - 2 e^{-W_1} - e^{- W_3 - W_4} \right)  \nn &+&
\frac{1}{20} \left( 3 b_1 e^{-W_1-W_4} + 3 b_2 e^{-W_1-W_3} - 2 a_1 e^{-W_2-W_4} - 2 a_2 e^{-W_2-W_3} \right), \nn &-& \frac{1}{40 g_7} (a_1 b_2 + a_2 b_1) e^{-W_1-W_2} - i\frac{3 g_7}{10 \sqrt{3}}(e^{-W_4} \slashed{c}_1 + e^{-W_3} \slashed{c}_2)  \nn &+& i\frac{3}{20} e^{W_2} \slashed{G}^2 - i \frac{1}{10} e^{W_1} \slashed{G}^1 \biggr] \e , \\
2 \g^{3} \delta  \psi_3 &=& e^{2 (\l_3+\l_4)} \biggl[ \frac{1}{10} \slashed{\partial}   ( W_1 + W_2 -2 W_3 +3 W_4) + \frac{g_7}{10} \left( 2 e^{-W_1} + 2 e^{-W_2} + e^{-W_3-W_4} \right) \nn &+& \frac{1}{20} \left( a_2 e^{-W_2-W_3} + b_2 e^{-W_1-W_3} -4 a_1 e^{-W_2-W_4} -4 b_1 e^{-W_1-W_4} \right) \nn &+& \frac{i}{20} e^{W_1} \slashed{G}^1+ \frac{i}{20} e^{W_2} \slashed{G}^2 -i\frac{3 g_7}{5 \sqrt{3}} e^{-W_3} \slashed{c}_2+i \frac{9 g_7}{10 \sqrt{3}} e^{-W_4} \slashed{c}_1- \frac{1}{20 g_7} (a_1 b_2 + a_2 b_1) e^{-W_1-W_2}  \biggr] \e, \nn
2 \g^{5} \delta  \psi_5 &=& e^{2 (\l_3+\l_4)} \biggl[ \frac{1}{10} \slashed{\partial } ( W_1 + W_2 +3 W_3 -2 W_4) + \frac{g_7}{10} \left( 2 e^{-W_1} + 2 e^{-W_2} + e^{-W_3-W_4} \right) \nn &+& \frac{1}{20} \left( a_1 e^{-W_2-W_4} + b_1 e^{-W_1-W_4} -4 a_2 e^{-W_2-W_3} -4 b_2 e^{-W_1-W_3} \right) \nn &+& \frac{i}{20} e^{W_1} \slashed{G}^1+ \frac{i}{20} e^{W_2} \slashed{G}^2 -i \frac{3 g_7}{5 \sqrt{3}} e^{-W_4} \slashed{c}_1+i \frac{9 g_7}{10 \sqrt{3}} e^{-W_3} \slashed{c}_2- \frac{1}{20 g_7} (a_1 b_2 + a_2 b_1) e^{-W_1-W_2}  \biggr] \e. \nonumber
\eea
In deriving these expressions, we have made use of (\ref{susy_cond}) and the inverse relations:
\bea
\l_1 &=& \frac{1}{10} (-3 W_1 +2 W_2 + W_3 + W_4), \quad \l_2 = \frac{1}{10} (2 W_1 -3 W_2 + W_3 + W_4), \nn
\l_3 &=& \frac{1}{10} ( W_1 + W_2 -2 W_3 +3 W_4), \quad \l_4 = \frac{1}{10}( W_1 + W_2 +3 W_3 -2 W_4).
\eea
We observe that the $\delta \chi_5$ supersymmetry variation offers nothing, since consistency demands that $\Gamma^i \chi_i = 0 \Rightarrow \Gamma^i \delta \chi_i = 0$. Taking various linear combinations, one finds the 3D algebraic supersymmetry variations:
\bea
\label{algebraic}
4 e^{-C - 2 \l_3 -2 \l_4} (\Gamma^1 \delta \chi_1+ \gamma^3 \delta \psi_3+ \gamma^5 \delta \psi_5) &=& \left[ \slashed{\partial } W_1 + \frac{i}{2} e^{W_1} \slashed{G}^1- 4 \partial_{W_1} T \right] \xi \otimes \eta_1 \otimes \eta_2, \nn
4 e^{-C - 2 \l_3 -2 \l_4} (\Gamma^2 \delta \chi_2+ \gamma^3 \delta \psi_3+ \gamma^5 \delta \psi_5) &=& \left[ \slashed{\partial } W_2 + \frac{i}{2} e^{W_2} \slashed{G}^2- 4 \partial_{W_2} T \right] \xi \otimes \eta_1 \otimes \eta_2, \nn
4 e^{-C - 2 \l_3 -2 \l_4} (-\Gamma^1 \delta \chi_1-\Gamma^3 \chi_3+ \gamma^5 \delta \psi_5) &=& \left[ \slashed{\partial } W_3 + \frac{i}{2}  e^{W_3} \slashed{F}^1- 4 \partial_{W_3} T \right] \xi \otimes \eta_1 \otimes \eta_2, \nn
4 e^{-C - 2 \l_3 -2 \l_4} (-\Gamma^1 \delta \chi_1-\Gamma^3 \chi_3+ \gamma^3 \delta \psi_3) &=& \left[ \slashed{\partial } W_4 + \frac{i}{2}  e^{W_4} \slashed{F}^2- 4 \partial_{W_4} T \right] \xi \otimes \eta_1 \otimes \eta_2,
\eea
where we have decomposed the spinor and dualised $c_I$ using (\ref{c_dual}). Following a procedure outlined in \cite{Colgain:2015mta}, we can also extract the following
\bea
\label{KSE}
&&e^{- (\l_3 + \l_4)} [\delta \psi_{a} + 2\gamma_{a} (\g^3 \delta \psi_3 + \g^5 \delta \psi_5 )] \nn &&= \biggl[ \mathcal{D}_{a} + T \rho_{a} + \frac{i}{8} \sum_{i=1}^2 \left( e^{W_i} \rho_{a}^{~bc}  G^i_{bc} + e^{W_{i+2}}  \rho_{a}^{~bc} F^{i}_{bc} \right) \biggr] \xi \otimes \eta_1 \otimes \eta_2,
\eea
where we have defined $\mathcal{D}_a \equiv \nabla_{a} - i \frac{g_7}{2} (B^1+B^2)_a$ and $C = -(\l_3+\l_4)$.

\end{document}